 \documentclass[10pt,preprint]{aastex}  
 \usepackage{natbib}
\def\ang{\AA}
\def\arcsec{\hbox{$^{\prime\prime}$}}

\def\gapprox{\lower.4ex\hbox{$\;\buildrel >\over{\scriptstyle\sim}\;$}}
\def\lapprox{\lower.4ex\hbox{$\;\buildrel <\over{\scriptstyle\sim}\;$}}

\shortauthors{ASCHWANDEN 2014}
\shorttitle{Loop-Guided Force-Free Fields}

\begin{document}

\title{         The Magnetic Field of Active Region 11158 During the 2011 February 12-17 Flares : 
		Differences between Photospheric Extrapolation and Coronal Forward-Fitting Methods }

\author{        Markus J. Aschwanden$^1$}

\affil{		$^1)$ Lockheed Martin Advanced Technology Center,
                Org. A021S, Bldg.252, 3251 Hanover St.,
                Palo Alto, CA 94304, USA;
                e-mail: aschwanden@lmsal.com }

\and

\author{        Xudong Sun$^2$ and Yang Liu$^2$
}

\affil{		$^2)$ W.~W.~Hansen Experimental Physics Laboratory,
		Stanford University, Stanford, CA 94305, USA;
		e-mail: xudongs@stanford.edu, yliu@sun.stanford.edu } 

\begin{abstract}
We developed a {\sl coronal non-linear force-free field (COR-NLFFF)} 
forward-fitting code that fits an approximate {\sl non-linear force-free 
field (NLFFF)} solution to the observed geometry of automatically traced 
coronal loops. In contrast to photospheric NLFFF codes, which calculate 
a magnetic field solution from the constraints of the transverse 
photospheric field, this new code uses coronal constraints instead, and 
this way provides important information on systematic errors of 
each magnetic field calculation method, as well as on the non-forcefreeness 
in the lower chromosphere. In this study we applied the COR-NLFFF code 
to active region NOAA 11158, during the time interval of 2011 Feb 12 to 17, 
which includes an X2.2 GOES-class flare plus 35 M and C-class flares. 
We calcuated the free magnetic energy with a 6-minute cadence over 
5 days. We find good agreement between the two types of codes for 
the total nonpotential $E_N$ and potential energy $E_P$, but find 
up to a factor of 4 discrepancy in the free energy $E_{free}=E_N-E_P$, 
and up to a factor of 10 discrepancy in the decrease of the free energy 
$\Delta E_{free}$ during flares. The coronal NLFFF code exhibits
a larger time variability, and yields
a decrease of free energy during the flare that is sufficient to satisfy
the flare energy budget, while the photospheric NLFFF code shows much
less time variability and an order of magnitude less free energy
decrease during flares. The discrepancy may partly be due to the 
pre-processing of photospheric vector data, but more likely due to
the non-forcefreeness in the lower chromosphere. We conclude that 
the coronal field cannot be correctly calculated based on photospheric 
data alone, but requires additional information on coronal loop
geometries.
\end{abstract}

\keywords{Sun: Corona --- Magnetic fields --- Sun: UV radiation}

\section{INTRODUCTION}

The art of computing solar magnetic fields has evolved from routine
potential field (PF) extrapolations to nonlinear force-free field (NLFFF)
computations. The most important aspect concerning these two types of
magnetic field models is the difference of magnetic energy content
between them, the so-called {\sl free energy}, which represents the
maximum energy that can be dissipated during a magnetic instability
that is driving solar flares, coronal mass ejections, or other eruptive
events. Monitoring the magnetic evolution in active regions became an
efficient diagnostic to measure the emergence of new magnetic flux
on the solar surface, the magnetic storage, the injection of electric 
currents and helicity, the step-wise dissipation of nonpotential energy 
during flares, the energy conversion of magnetic energy
during a magnetic reconnection process into particle acceleration, etc.
The quantification of the free magnetic energy probably provides the strongest
constraint to discriminate cause and consequence in the various 
energy conversion processes. It is therefore imperative to develop 
accurate and efficient tools to measure and monitor the evolution of 
magnetic energy in solar active regions. In this study we explore 
a novel method to calculate an approximate NLFFF solution using an automated 
loop tracing code, which aims to find force-free magnetic field solutions 
that are most consistent with the observed geometry of coronal loops, 
such as observed in high-resolution extreme-ultraviolet (EUV) images from the 
{\sl Atmospheric Imaging Assembly (AIA)} (Lemen et al.~2012) onboard the 
{\sl Solar Dynamics Observatory (SDO)} spacecraft (Pesnell et al.~2012).

NLFFF Modeling has been applied to active regions (Bobra et al.~2008;
Su et al.~2009a; Inoue et al.~2011, 2013; Savcheva et al.~2012a), 
to coronal sigmoids (Savcheva et al.~2012b, 2012c; Inoue et al.~2012),
to photospheric magnetic field changes (Liu et al.~2012),
to magnetic flux emergence and energy build-up before flares (Su et al.~2009b;
Li et al.~2007), to flares (Guo et al.~2008; Schrijver et al.~2008), 
to confined eruptions (Guo et al.~2010), to Coronal Mass Ejections (CMEs) 
(Su et al.~2011; Feng et al.~2013),
as well as to MHD simulations (Savcheva et al.~2012a).
Earlier studies compared theoretical models with observed images in a
rather qualitative way, but recent studies go into quantitative comparisons
of free energies and statistics of misalignment angles between modeled
field lines and observed coronal loop geometries.
Comparisons of up to 14 different NLFFF codes revealed substantial
differences in the amount of calculated free magnetic energies,
ranging from an (unphysical) minumun value $E_{N}/E_P=0.88$ below unity,
to a maximum of $E_{N}/E_P=1.30$ for an X3.4-class flare
 (Schrijver et al.~2008). 
An evaluation of current NLFFF modeling was conducted in a study of the
solar active region NOAA 10953 (2007 April 30) with a dozen of different NLFFF
codes, which led to three major requirements for successful NLFFF 
modeling: (1) a sufficiently large field-of-view of the vector magnetic 
field data and the computation box, (2) accomodation of uncertainties 
in the boundary data, and (3) a realistic model of the transition from 
the non-forcefree photosphere to the force-free base of the corona 
(DeRosa et al.~2009). A pre-processing step to minimize the net force and
torque in the photospheric boundary data was suggested 
(Wiegelmann et al.~2006, 2008; Wheatland and Regnier 2009), which improved
the extrapolation above the chromosphere, but the field connectivity 
and free magnetic energy were not well recovered (Metcalf et al.~2008;
Yamamoto et al.~2012; Jiang and Feng 2013).
The latter issue is manifested in a major discrepancy between the 
computed magnetic field lines and stereoscopically observed geometries 
of coronal loops (Fig.~1), amounting to 3D misalignment angles of $\approx 
24^\circ-44^\circ$ (DeRosa et al.~2009). This mismatch angle can be
reduced by forward-fitting of parameterized potential field models
(Aschwanden and Sandman 2010; Sandman and Aschwanden 2011), 
by forward-fitting using a quasi-Grad-Rubin NLFFF method (Malanushenko 
et al.~2011, 2012), and by forward-fitting of analytical NLFFF 
approximations (Aschwanden et al.~2012, Aschwanden 2013a, 
2013b, 2013c; Aschwanden and Malanushenko 2013). 
The latter method has been 
demonstrated to work equally well for both 3-dimensional (3D) loop 
coordinates, obtained with stereoscopy, and 2-dimensional (2D) 
loop projections (Aschwanden 2013c).

In this study we developed our forward-fitting code with analytical
NLFFF approximations further by (1) implementing an automated loop 
tracing code for detection of coronal loops in multi-wavelength
EUV images, which makes the manual or visual loop tracing unnecessary 
(Aschwanden 2010), (2) by optimization of the forward-fitting
technique to 2D loop coordinates, which relinquishes 3D reconstruction
with stereoscopy (Aschwanden 2013c), and (3) by improved detection
and rejection of false loop structures, caused by CCD pixel bleeding,
CCD saturation, and diffraction patterns from by the EUV telescope
entrance filter. The content of this paper is
a brief analytical description of the method (Section 2), a brief
description of the numerical code (Section 3), and data analysis 
of NOAA active region 11158 observed with SDO/AIA and HMI during 
the 2011 February 12-17 flares (Section 4), discussion and comparison with
previous studies on the same active region and X-class flare (Section 5), 
and conclusions (Section 6).

\section{ANALYTICAL DESCRIPTION OF MAGNETIC MODELING}

Our {\sl Coronal Nonlinear Force-Free Field (COR-NLFFF)} forward-fitting code 
consists of the following three major tasks (Fig.~2): (1) Automated loop
tracing in coronal 2D images; (2) decomposition of a line-of-sight magnetogram
into (sub-photospheric) magnetic charges that provide a parameterization 
of the potential and non-potential field; and (3) forward-fitting of the 
nonlinear force-free field approximation to the observed, automatically
traced 2D loop coordinates. A brief theoretical description of these three 
modeling steps is provided in the following. 

\subsection{Automated Loop Tracing}

An early version of the automated loop tracing code, called {\sl Oriented
Coronal CUrved Loop Tracing (OCCULT)}, has been quantitatively compared with 
four other codes (Aschwanden et al.~2008). The method is based on
oriented-directivity tracing of curvi-linear features, but in contrast to
other general feature-extraction algorithms, it is customized to solar EUV
and SXR images by taking advantage of the specific property that coronal 
loops have large curvature radii, compared with their widths. 
Essentially, an image is highpass-filtered to enhance curvi-linear features
with a small width, which are then traced with a guiding criterion that is
defined in terms of the local curvature radius within small 
directional changes along a traced loop segment. The performance of this
code was systematically improved by optimizing the guiding criterion, 
using the first-order term of the loop direction in the first version 
(Aschwanden 2010), and using the second-order term of the local curvature
radius in the latest version OCCULT-2 (Aschwanden, De Pontieu, and Katrukha 
2013b). 

\subsection{Potential Field Computation}

For the computation of a potential field, a line-of-sight (LOS) 
magnetogram that samples one magnetic field vector component $B_z(x,y)$
is sufficient to calculate a 3D potential field in a given computation
box. While standard potential-field codes, such as the widely used 
{\sl Potential Field Source Surface (PFSS)} code, are based on the 
{\sl eigen function (spherical harmonic) expansion} method originally 
developed by
Altschuler and Newkirk (1969), we need a parameterization that can also
be used for forward-fitting of non-potential 3D magnetic field models.
The simplest method that is suitable for this purpose is the decomposition
of a potential field into uni-polar magnetic charges $j=1,...,N_m$ 
that are buried in sub-photospheric locations ($x_j, y_j, z_j$), with 
the field strength $B(r)$ decreasing with the square of the radial 
distance $r$, which can be superimposed by an arbitrary large number 
$N_m$ of magnetic charges,
\begin{equation}
        {\bf B}({\bf x}) = \sum_{j=1}^{N_{\rm m}} {\bf B}_j({\bf x})
        = \sum_{j=1}^{N_{\rm m}}  B_j
        \left({d_j \over r_j}\right)^2 {{\bf r_j} \over r_j} \ ,
\end{equation}
where $r_j=[(x-x_j)^2+(y-y_j)^2+(z-z_j)^2]^{1/2}$ is the distance of an
arbitrary coronal location ${\bf x}=(x,y,z)$ to the subphotospheric charge
location $(x_j, y_j, z_j)$, while $d_j=1-[x_j^2+y_j^2+z_j^2]^{1/2}$ is the
depth of the buried charge, and $B_j$ is the magnetic field strength
at the solar surface in vertical direction above the buried charge.
The square-dependence of the radial field component $B(r) \propto r^{-2}$ 
warrants that each magnetic charge fulfills Maxwell's divergence-free 
condition,
\begin{equation}
	\nabla \cdot {\bf B} = 0 \ ,
\end{equation}
which it is also true for the summed magnetic field according to Equation (1),
because the linear superposition of divergence-free fields is 
divergence-free too, i.e., $\nabla \cdot {\bf B} = \nabla \cdot 
(\sum_j {\bf B_j}) = \sum_j (\nabla \cdot {\bf B_j}) = 0$. 

The decomposition of a LOS magnetogram $B_z(x,y)$ into a finite
number $N_m$ of magnetic charges can simply be accomplished by iterative 
decomposition of local maxima of the field into individual magnetic charges,
each one characterized by the four parameters $(B_j, x_j, y_j, z_j),
j=1,..., N_m$, by iterating subsequent subtractions of decomposed
Gaussian-like magnetic charges until the residual map reaches some
noise threshold. The procedure is demonstrated in detail in Aschwanden
and Sandman (2010). Typically, a number of $N_m \approx 100$ magnetic
sources is sufficient to obtain a realistic potential field model of
a solar active region.

\subsection{Nonlinear Force-Free Field Computation}

A {\sl nonlinear force-free field (NLFFF)} model needs to fulfill
both the Maxwell's divergence-free (Equation 2), and the force-free condition,
\begin{equation}
         {{\bf j} \over c} = {1 \over 4\pi} (\nabla \times {\bf B}) =
         \alpha({\bf x}) {\bf B} \ .
\end{equation}
where $\alpha({\bf x})$ is a scalar function that varies in space,
but is constant along a given field line, and the current density ${\bf j}$
is co-aligned and proportional to the magnetic field ${\bf B}$.
An approximate analytical solution of Equations (2)-(3) was recently
calculated for vertically twisted fields (Aschwanden {\bf 2013a}) that can be 
expressed by a superposition of an arbitrary number of $N_m$ magnetic 
field components ${\bf B}_j$, $j=1,...,N_m$,
\begin{equation}
        {\bf B}({\bf x}) = \sum_{m=1}^{N_m} {\bf B}_j({\bf x}) \ ,
\end{equation}
where each magnetic field component ${\bf B}_j$ can be decomposed
into a radial $B_r$ and an azimuthal field component $B_\varphi$,
\begin{equation}
        B_r(r, \theta) = B_m \left({d^2 \over r^2}\right)
        {1 \over (1 + b^2 r^2 \sin^2{\theta})} \ ,
\end{equation}
\begin{equation}
        B_\varphi(r, \theta) =
        B_m \left({d^2 \over r^2}\right)
        {b r \sin{\theta} \over (1 + b^2 r^2 \sin^2{\theta})} \ ,
\end{equation}
\begin{equation}
        B_\theta(r, \theta) \approx 0
        \ ,
\end{equation}
\begin{equation}
        \alpha(r, \theta) \approx {2 b \cos{\theta} \over
        (1 + b^2 r^2 \sin^2{\theta})}  \ ,
\end{equation}
where ($r, \varphi, \theta$) are the spherical coordinates of a
magnetic field component system ($B_j, x_j, y_j, z_j, \alpha_j)$
with a unipolar magnetic charge $B_j$ that is buried at position
($x_j, y_j, z_j)$, has a depth $d=1-(x_j^2+y_j^2+z_j^2)^{1/2}$,
a vertical twist $\alpha_j$, and $r=[(x-x_j)^2+(y-y_j)^2+(z-z_j)^2]^{1/2}$
is the distance of an arbitrary coronal position $(x,y,z)$ to the
subphotospheric location $(x_j, y_j, z_j)$ of the buried magnetic charge.
The force-free $\alpha$ parameter can be expressed in terms
of the parameter $b$ (Equation 8), which quantifies the number $N_{twist}$
of full twist turns over a (loop) length $L$,
\begin{equation}
        b = {2 \pi N_{twist} \over L} .
\end{equation}
This analytical approximation is divergence-free and force-free
to second-order accuracy in the parameter $(b\ r \sin \theta)$
(Aschwanden 2013a), which is proportional to the force-free parameter 
$\alpha$ as defined by Equation (8). In the limit of
vanishing vertical twist ($\alpha \mapsto 0$ or $b \mapsto 0$),
the azimuthal component vanishes, $B_\varphi \mapsto 0$, and the
radial component degenerates to the potential-field solution of
a unipolar magnetic  charge, $B_r \mapsto B_j (d/r)^2$, which
is simply a radial field that points away from the buried charge
and decreases with the square of the distance. As a caveat, we should
be aware that the analytical approximation expressed in Equations
(5)-(6) implies a vertical twist axis that produces
a horizontal azimuthal non-potential field component $B_\varphi$,
while horizontal twist axes with the corresponding 
non-potential components $B_\varphi$ cannot be represented with
this parameterization, such as a horizontally oriented filament
(see also discussion on the Gold-Hoyle flux rope in Appendix A
of Aschwanden 2013a).  We will label the nonpotential magnetic field 
of our model with a vertical twist axis as $B_{N,\perp}$, in order 
to remind about this restriction.

\subsection{Computation of Free Magnetic Energy}

The free energy is generally
defined as the difference between the potential and non-potential
magnetic energy, integrated over the volume of the computation box.
In our case, since we include only vertical twist axes in our nonpotential
model, the difference energy $E_{\perp}^{free}$ will be a lower limit
to the total free energy $E^{free}$,
\begin{equation}
	E_{\perp}^{free} = E_{N,\perp} - E_P =
	{1 \over 8 \pi} \left( \int B_{N,\perp}^2({\bf x}) dV - 
	                       \int B_{P}^2({\bf x}) dV \right) \ .
\end{equation} 
In order to obtain a first-order correction, we can consider a
current in a semi-circular loop or filament. This current flows in
vertical direction at the footpoint of a loop or filament, while it
flows in horizontal direction near the apex. This twist axis of the
field-aligned current $j_\parallel$ has thus a relative direction $\theta$
with a cosine-dependence along the loop, which will also enter the resulting 
nonpotential azimuthal field component $B_{\varphi}$ and leads to an average 
underestimate (as a function of the inclination angle $\theta$ to
the vertical) of 
\begin{equation}
	\langle B_{\varphi,\perp} \rangle 
	= \langle B_\varphi \cos(\theta) \rangle
	= \langle B_\varphi \rangle \left({2 \over \pi}\right) \ .
\end{equation}
The free energy $E_{free}$ scales with the square of the azimuthal 
(nonpotential) field component $B_{\varphi}$ by definition (Equation 10 
in Aschwanden 2013a), and thus the average free energy $E_{\perp}^{free}$ 
obtained in a model that only includes electric currents associated 
with a vertical axis is
\begin{equation}
	 E^{free}_{\perp}     =  E^{free} \langle \cos(\theta)^{-2} \rangle
	                    =  E^{free}  \left({2 \over \pi}\right)^2
			    =  E^{free} / q_{iso} \ .
\end{equation}
Thus we expect that our method underestimates the azimuthal (non-potential)
field component $B_{\varphi}$ by a factor of $(\pi/2) \approx 1.6$ in the 
average, and thus the free energy by a factor of $(\pi/2)^2 \approx 2.5$. 
This bias can approximately be corrected with the {\sl isotropic twist 
correction factor} $q_{iso}=(\pi/2)^2 \approx 2.5$ in the determination
of the free energy, i.e., $E^{free}=E^{free}_{\perp} q_{iso}$.

\section{NUMERIC CODE}

The {\sl Coronal Nonlinear Force-Free Field (COR-NLFFF)} forward-fitting 
numerical code has evolved from an earlier version (Aschwanden and
Malanushenko 2013) to the present version, achieving now a higher degree
of accuracy in the measurement of the nonpotential magnetic 
energy $E_{N}$ and the free energy $E_{free}=E_{N}-E_P$. 
A flow chart of the latest version is provided
in Fig.~2, which can be broken down into three major tasks: (1) the
processing of the magnetic data; (2) the processing of the EUV data
(including automated loop tracing), and (3) the forward-fitting of the
analytical NLFFF approximation. In the following we briefly describe 
the numerical methods that are employed in the COR-NLFFF code, while
a parametric study to establish the optimum choice of control parameters 
(Table 1) is documented in Appendix A. 

\subsection{The Magnetic Field Decomposition}

The input of magnetic data is a line-of-sight magnetogram $B_z(x,y)$, 
in this case from HMI (Scherrer et al.~2012; Hoeksema et al.~2014) 
onboard the SDO spacecraft. We use the 45 s HMI data, which have been
processed from 135 s time intervals. 
An HMI image covers the full-Sun and has a pixel size 
of $0.5\arcsec$, or approximately $\Delta x=0.00052$ solar radii. 
We are not using any photospheric vector magnetograph data, in contrast to the 
traditional NLFFF codes, because we consider the transverse components 
$B_x(x,y)$ and $B_y(x,y)$ as unreliable for coronal field extrapolations 
due to the non-forcefreeness of the photosphere (Fig.~1), and thus we 
keep them as free parameters, which will be computed from the constraints 
of the geometric shapes of observed coronal loops. We use only the
line-of-sight magnetic field component $B_z(x,y)$ of the HMI data.
The magnetic control parameters of the COR-NLFFF code are: the spatial 
resolution ($\Delta x_{mag}$) of (rebinned) magnetogram data used in 
the decomposition of (gaussian-like) point sources, and the number of 
decomposed magnetic source components ($n_{mag}$). The numerical 
algorithm for magnetogram decomposition is described in Aschwanden 
(2010) and Aschwanden et al.~(2012, Appendix A therein). A modification
in the newest version of the code is the fitting of a 2D Gaussian
function to local maxima of the magnetograms, which are iteraticely 
subtracted and the residuals within the full width at half mean (FWHM)
are set to zero, in order to minimize the number of magnetic source
components in the model. In addition, we normalize the fitted Gaussian 
to the same magnetic field strength as the local peak in the magnetogram,
in order to conserve the magnetic flux ($\Phi = B \times {\rm FWHM}^2$)
and magnetic energy ($E_B \propto B^2$). Examples of the Gaussian
decomposition of the magnetogram are given in the parameteric study 
described in Appendix A (see also Figure 17). 

\subsection{The Selection of Loops}

The second major task deals with the extraction of coronal loop coordinates
from the EUV images. The optimization of the automated loop tracing code
is described in a recent study (Aschwanden et al.~2013b), 
which is controlled by the lowpass filter constant ($n_{sm1}$) 
(boxcar pixels), the highpass filter ($n_{sm2}=n_{sm1}+2$), the minimum
loop curvature radius ($r_{min}$), the minimum loop length
($l_{min})$, and the loop length increment ($\Delta s$), which
all have been optimized (Appendix A) and the default values 
are given in Table 1.

A unique capability of our COR-NLFFF code is the synthesized processing of 
multiple EUV images in different wavelength filters, which encompsasses
six coronal (94, 131, 171, 193, 211, 335 \ang ) and one chromospheric
(304 \ang ) filter from AIA/SDO. How do we optimize the selection of 
loops in the forward-fitting code?  We set up a number of control 
parameters that include the selected wavlengths ($\lambda$), the maximum 
number ($n_{loop}$) of loops extracted per wavelength filter, the 
minimum loop length ($l_{min}$), and the minimum curvature radius
($r_{min}$) in the loop selection. 
We find that selections with an over-proportional contribution of large
loops tends to favor potential field solutions (and thus a small
ratio of the non-potential energy $q_E=E_{N}/E_P$, while a selection
with a dominant contribution of small loops produces the opposite.
Since large loops are found at relatively cool temperatures $(T \approx
1-2$ MK; i.e., 171 and 193 \ang ), while smaller loops in the core of
active regions are found at hotter temperatures $(T \approx 10-20$ MK;
i.e., 94 and 335 \ang ), the choice of wavelengths ($\lambda$) and the
limit $(n_{loop}$) on the extracted loops per wavelength filter
play a decisive role. It is crucial to have a balanced amount of small 
and large loops in order to obtain a representative value for the free 
energy. A parametric study of varying the selection parameters 
($\lambda$, $n_{loop}$, $l_{min}$, $r_{min}$) is given in Appendix A.

We implemented also criteria to automatically reject false loop
structures, i.e., curvi-linear features that are caused by CCD
pixel bleeding, CCD saturation, and diffraction patterns from the EUV
entrance filter. 

\subsection{ Forward-Fitting Control Parameters }

The forward-fitting algorithm 
consists of the task of optimizing the free parameters $\alpha_j,
j=1,...,N_m$, associated with each magnetic charge in such a way
that the 2D misalignment angle $\mu_2$ between an observed loop
segment and a theoretical field line calculated at the same location 
is minimized. Forward-fitting with a large number of free parameters
requires a customized numerical scheme that makes a reasonable
trade-off between numerical accuracy and computational efficiency.
A first numerical code that fits the analytical approximation
(Equations 4-9) to a given analytical 3D magnetic field is described 
and tested in Aschwanden and Malanushenko (2013), forward-fitting
to real solar data in 2D and 3D is conducted in Aschwanden (2013c)
and computations of the free energies in Aschwanden (2013b).
The multi-parameter optimization scheme used in the original code
(Aschwanden and Malanushenko 2013) employed a hierarchical
subidivision of $\alpha$-zones in the magnetic map that refines
iteratively the $\alpha$-values into progressively smaller regions.
In the latest version we use all magnetic source components 
in the forward-fitting of force-free $\alpha_j$ parameters.
In addition we tested different optimization algorithms: (1) the Powell 
optimization scheme (which seeks the local minima of the parameters 
$\alpha_j$ sequentially during each iteration cycle), and (2) a gradient 
optimization scheme (which determines the gradient $d\mu_j/d\alpha_j$ 
of the misalignment angle $\mu_j$ as a function of the free parameter 
$\alpha_j$ during each iteration cycle. The Powell optimization generally 
converges to a smaller misaligment angle and is the method of choice 
in the COR-NLFFF code. 

The mean misalignment angle $\mu_2$, which is minimized in the forward-fitting
method, is defined by the root-mean-square (r.m.s.),
\begin{equation}
	\langle \mu_2 \rangle = \sqrt{ {\sum_i \mu_i^2 \over n_{loop}} }
	\ , \qquad i=1,...,n_{loop} \ ,
\end{equation}
where $\mu_i$ is the misalignment angle of loop $i$, which itself is
the median of $n_{seg}$ loop segments (typically $n_{segm}=7$),
\begin{equation}
	\mu_i = median(\mu_{i,k}) 
	\ , \qquad k=1,...,n_{segm} \ .
\end{equation}
Thus, the forward-fitting essentially consists of fitting the  
$\alpha$-values $(\alpha_j, j=1,...,n_{fit})$ to the 
loop misalignment angles $\mu_{i,k}$, with $i=1,...,n_{loop}$ 
and $k=1,..., n_{segm}$. The optimization of the forward-fitting
control parameters ($n_{seg}, \Delta \alpha_j, h_{max}, n_{iter}$)
is quantified in Appendix A.

There is also the problem of the missing third dimension in the observations.
The forward-fitting algorithm minimizes the misalignment angle $\mu_2$
between an observed 2D loop coordinate $(x_i, y_i), i=0,...,n_{fit}$
and a theoretical 3D field line intersecting at $(x_i, y_i, z_i)$. 
The line-of-sight coordinate $z_i$ is not known a priori, unless we
would employ stereoscopy (Aschwanden 2013c), but is a required quantity
to select the proper theoretical field line. One constraint that we can
use is the valid range of $z_i=[(1+h_i)^2-x_i^2-y_i^2]^{1/2}$  
within a solution space corresponding to an altitude range of 
$0 < h_i < h_{max}$ at each position $(x_i,y_i)$. We assume that each 
observed loop segment can be approximated with a circular 
segment as a function of the LOS-coordinate $z_i$,
\begin{equation}
	z_{i,k}(s) = z_{i,0} + r_{curv} \sin[\varphi_1 
	+ (\varphi_2-\varphi_1) q_{seg,k}]
	\ ,
\end{equation}
where the starting angle $\varphi_1$ of the circular segment
can be anywhere within the range of $0^\circ \le \varphi_1 \le +180^\circ$, 
while the end angle $\varphi_2$ can be anywhere within the range of 
$\varphi_1 \le \varphi_2 \le +180^\circ$, where 
$0 < q_{seg} < 1$ is a normalized loop length coordinate,
$r_{curv}$ is the curvature radius of the circular segment,
and $z_0$ is the center height of the circular segment, adjusted in such a way
that the circular segment is always located in the altitude range of
the computation box with $h=[0, h_{max}]$. We choose $n_c=n_{seg}=7$ starting
angles $\varphi_1$ (each $30^0$ apart) and $n_h=n_{seg}=7$
possible height ranges, which yields $n=n_h \times
n_c (n_c-1)/2 = 147$ different trial geometric shapes to estimate the
heights $z_{i,k}$ in the $k=1,...,n_{seg}$ loop segments of each loop $i$. 

After every iteration cycle of forward-fitting a set of $\alpha_j$-values, 
we compute the trial geometric shapes (Equation 15) with the new $\alpha_j$ 
values and adjust 
the heights $z_i^{trial}$ at the midpoints of each loop, by minimizing the
misaligment angles $\mu_2$ between the loop coordinates
($x_i, y_i, z_i^{trial}$) and the theoretical field lines
($x_i, y_i, z_i^{model}$). In this way we optimize both the
force-free parameters $\alpha_j$ and the 3D-coordinate $z_i$ of the
loop midpoints. Alternative trial geometries used for magnetic modeling
of 3D loops are Bezier functions (Gary et al.~2014a,b).

\section{OBSERVATIONS AND DATA ANALYSIS}

SDO data are most suitable for our type of magnetic modeling of active 
regions, because HMI and AIA provide both high-resolution magnetograms
and simultaneous multi-wavelength EUV images. An additional requirement for a 
suitable active region is a location near Sun disk center and the appearance
of many (possibly twisted) coronal loops. Some of such active regions
have already been modeled with NLFFF codes that allow us to compare results. 
The active region NOAA 11158 produced one of the most spectacular flares
(GOES X2.2 class) during the first two  years of the SDO mission and has been
intensively studied since then, documented in over 40 publications so far
(e.g., Sun et al.~2012a; Schrijver et al.~2011; Wang et al.~2012;
Kosovichev 2011, Aschwanden et al.~2013a). See Section 5.1 for more details
and references.

\subsection{AIA and HMI Observations}

We study the evolution of active region 11158 over 5 days, from 2011 
February 12, 00 UT, to February 17, 00 UT, which is the same
interval as analyzed in Sun et al.~(2012a). A sample set of 7 images,
observed with AIA/SDO (Lemen et al.~2012; Boerner et al.~2012) on 2011 
February 15, 00:00:00 UT (about two hours before the GOES X2.2-class flare),
is shown in Fig.~3 with the wavelengths of 94, 131, 171, 193, 211, 304,
and 335 \ang\ . A near-simultaneous magnetogram observed with HMI/SDO 
within one minute (2012 February 14, 23:58:57 UT) is shown in
Fig.~3 also. The pixel size of AIA is $0.6\arcsec$ (with a spatial 
resolution of
$\approx 2.5$ pixels), and the pixel size of HMI is $0.5\arcsec$.
These 8 images shown in Fig.~3 represent all data input that is needed
for our magnetic NLFFF modeling. No vector magnetograph data from HMI 
are used, because our method needs only the line-of-sight component
and treats the transverse magnetic field components as free parameters.

We show the loop structures enhanced in form of bandpass-filtered images 
in Fig.~4. The bandbass used here consists of a lowpass filter
with a boxcar length of $nsm_1=5$ pixels, and a highpass filter with a
boxcar length of $nsm_2=7$ pixels. Thus it enhances most efficiently
loops with a cross-sectional width in the range of 5-7 pixels, which
corresponds to a cross-sectional width of $w\approx 2-3$ Mm. 
Obviously we see coronal loops in
all wavelengths (Fig.~4), but filters with cool coronal temperatures 
(171, 193, 211 \ang ; $T_e\approx 1-2$ MK) show more large-scale loops that 
overarch the active region, while filters sensitive to hotter temperatures
(94, 131 \ang) reveal more highly twisted structures in the core of the
active region, similar to the structures seen at chromospheric temperatures
(304 \ang ; $T \approx 0.05$ MK). Loop structures in this active region
were analyzed in almost the same set of images two hours before the X-class 
flare, and have been subjected to automated temperature and emission 
measure analysis and automated loop tracing (Aschwanden et al.~2013a).

\subsection{Automated Loop Tracing}

The automated loop tracing in this active region is shown separately
for each temperature filter in Fig~5, conducted with an improved
algorithm (Aschwanden et al.~2013a) than used previously. 
For the particular
run shown in Fig.~5 we used the following control parameters (defined
in Aschwanden et al.~2012):
a maximum number of $n_L=50$ loops per wavelength filter,
a minimum curvature radius of $r_{min}=30$ pixels, a minimum loop length
of $l_{min}=30$ pixels, a lowpass filter of $n_{sm1}=5$ pixels, 
a highpass of $n_{sm2}=7$ pixels, and a flux threshold level that 
corresponds to [5, 2, 0, 0, 1, 5, 5] times the noise level (median)
for the wavelengths [94, 131, 171, 193, 211, 304, 335] A,  
evaluated in stripes at each of the four image boundaries.
With this particular setup the code picked up a total of $n=238$ 
curvi-linear loop structures (Fig.~5, bottom right), synthesized from
all wavelengths, from a minimum of $n=89$ structures in the 94 \ang\
filter, and a maximum of $n=149$ structures in the 193 \ang\ filter.
The automated tracing reveals a large number of bipolar loops and peripheral
(open-field) fan structures in the 171, 193, and 211 filters, while 
the other filters exhibit shorter structures in the core of the active 
region surrounding the flare site. The complementary nature of loop
tracing is most conspicuous between 171 and 335 \ang\ (Fig.~5).

\subsection{Nonlinear Force-Free Field Computation}

The results of the forward-fitting of the analytical NLFFF approximation 
is shown for each wavelength separately in Fig.~6. The panels in
Fig.~6 show the automatically traced 2D coordinates of the loop segments 
(blue curves) and their midpoints (blue diamonds), where an intersecting
fieldline segment of the NLFFF best-fit model with equal length is 
indicated (red curve), as well as the full field line (orange curve), 
starting and
ending either at the solar surface or at a boundary side of the
computation box, which extends over the field-of-view shown in the
Figures 3-6 and a vertical height of $h_{max}=0.15$ solar radii above the
solar surface. The control parameter of these fits are:
$n_{mag}=100$ magnetic sources; a spatial resolution corresponding to 
one AIA pixel size; a maximum range of the force-free parameter
$\Delta \alpha_{max}=\pm 10 R_{\odot}^{-1}$ per iteration step; a maximum 
number of $n_{loop}=50$ loops per wavelength filter; and a minimum length of 
$l_{min}=30$ pixels for the selected loops. A simultaneous 
fit to a subset of 238 loops synthsized from the 50 brightest ones
in each filter is shown in Fig.~6 (bottom right panel).

It is interesting to compare the best-fit NLFFF solutions in each 
wavelength, because they reveal the model sensitivity to the 
loop selection. The wavelengths with the largest number of loops
are 171, 193, and 211 \ang , amounting to $n_{loop} \approx 110-150$. 
The other wavelengths, 
however, have a smaller number of loops ($n_{Loop}\approx 10-70$), have
smaller loop lengths, are preferentially located in the core of the 
active region, and have lower altitudes. It appears that these two
groups of loops have also different degrees of non-potentiality:
the large-scale loops over-arching the active region are close to
the potential field solution with low $\alpha$-values, while the
small-scale loops in the core of the active region are stronger
twisted, including the sigmoid-like filament in the center, and have 
a higher degree of non-potentiality, with higher $\alpha$-values. 

The relaxation of a twisted non-potential field to a more dipolar
potential field can also be seen visually by inspecting two of the
COR-NLFFF solutions for AR 11158 before the X2.2 flare (Fig.~7, top
panel: 2011-Feb-15, 01:48 UT), and shortly after the flare peak (Fig.~7, bottom
panel: 2011-Feb-15, 02:06 UT). The highly twisted central filament in 
the northern pair of sunspots of the quadrupolar active region 
relaxes from a helical twist of about one turn to an almost
dipolar geometry with little twist. A movie of the magnetic field
caclulated with COR-NLFFF during a time interval of 6 hours 
is contained in the supplementary electronic material to this paper.

\subsection{Time Evolution of the Free Energy}

We study now the time evolution of the potential magnetic 
energy $E_P(t)$, the non-potential magnetic energy $E_{N}(t)$, and the
free magnetic energy or difference $E_{free}(t)=E_{N}(t)-E_P(t)$
for AR 11158.
These energies are integrated over a computation box that is aligned with 
the spherical solar surface, with a field-of-view of $0.3 \times 0.3$
solar radii, and with an altitude range of $h_{max} = 0.15$ solar radii, 
centered at heliographic coordinates $[l(t),b(t)]$ following the
solar rotation. The central meridian passage of AR 11158 was at 
2011 Feb 14, 00 UT, and the heliographic latitude is $b(t) \approx -15^\circ$.

The evolution of the potential magnetic energy starts with a low amount
of $E_P\approx 1 \times 10^{32}$ erg on 2011-Febr-12, and then ramps up
almost monotonically to a value of $E_P \approx 8 \times
10^{32}$ erg during the next two days (Fig.~8b), so the total magnetic
energy increases by a factor of about 8 during the 2 days before the X2.2
flare on 2011-Feb-15. Both our COR-NLFFF code and the Wiegelmann NLFFF code 
(Sun et al.~2012a) agree well (within a few percents) in the total potential 
energy during the first 3 days, but deviate by about 
30\% during the last two days, when the active region was $\gapprox 25^\circ$ 
away from the centeral meridian, perhaps due to some side effect of 
the de-rotation projection to disk center applied in Sun et al.~(2012a), 
or due to center-of-limb effects of the LOS-component decomposition in our
COR-NLFFF code. 

The time evolution of the nonpotential magnetic energy $E_N(t)$, obtained from 
$2 \times 1200$ forward-fits with the COR-NLFFF code to traced coronal loops, 
with a 6-min cadence, is shown in Fig.~8c (blue profile), 
and is compared with the Wiegelmann NLFFF code with 12-min intervals (red curves), 
calculated previously also wtih 1-hour intervals (orange curves in Fig.~8; 
see also Fig.~4 in Sun et al.~2012a). 

We show the evolution of the free energy $E_{free}(t)=E_N(t)-E_P(t)$ 
in Fig.~(8d). There is a rapid increase in the free energy 
at the beginning of the second day (2011-Feb-13) as calculated with the 
Wiegelmann code, while no such jump is evident as caclulated from the HMI
line-of-sight component with the COR-NLFFF code, which is interpreted as
an episode of rapid magentic flux emergence with strong horizontal field
and little vertical field, occurring in the north-western bipole of the active 
region. Apparently there are no bright loops associated with the emerging field,
because we detect no increase in free energy based on the coronal NLFFF code.

Since the free energy is a small quantity obtained from the
difference of two large quantities, the relative uncertainties are 
larger than the relative uncertainties of the absolute energies. We do not 
know {\sl a priori} which code produces more reliable values of the free
energy, since each code has its own caveats and strengths, and the two
codes are using complementary (photospheric versus coronal) information.
We find the following differences in the results of the free energy 
obtained with the Wiegelmann NLFFF code, which uses pre-processed 
photospheric vector data (Sun et al.~2012a), and our COR-NLFFF code, 
based on forward-fitting to (automatically) traced coronal loops: 
(1) The absolute values of the free energies differ up to a
factor of 4 between the two codes; (2) The Wiegelmann NLFFF code 
yields a relatively smooth evolution of the magnetic energy, while our 
COR-NLFFF reveals more rapid fluctuations in the time evolution of 
the free energy (Fig.~8d); 
(3) There are a number of GOES flares where the COR-NLFFF code shows 
substantial changes in the free energy that appear to be correlated 
with the GOES fluxes, while the Wiegelmann NLFFF code shows only very 
small fluctuations of the free energy. We will investigate these differences
in more detail in the following. 

Let us have a look at some fitting parameters in order to see whether 
they could be responsible for the fluctuations of the obtained free energy. 
The number of automatically selected loops in the 1200 forward-fitting runs 
are shown in Fig.~(8e), and has a mean of $n_{loop}=248 \pm 18$. 
The number of selected loops has a lowest value of $n_{loop} = 192$
during the first day, when
the GOES flux as well as the total magnetic energy is also low, which
reflects the fact that less detectable loops existed during this day.
In the overall, there is much less fluctuations in the number of loops
$n_{loop}(t)$ than in the free energy $E_{free}(t)$ (Fig.~8d).

The goodness-of-fit test of the 1200 forward-fitting runs is given in
terms of the median 2D misalignment angle $\mu_2(t)$ in Fig.~(8f),
with a mean of $\mu_2=4.4^\circ\pm0.7^\circ$, 
which is to be compared with $\mu_2 \approx 10^\circ$ for potential 
field models. The goodness-of-fit $\mu_2(t)$ is fairly uniform
during the 5 days, even during flaring episodes, thanks to the new
improvements in the COR-NLFFF code that eliminates false loop structures 
caused by CCD pixel bleeding, CCD saturation (for too long 
exposure times), and diffraction patterns due to the EUV entrance filters.

\subsection{Free Energy Changes and Correlations with GOES Flares}

We show an expanded time profile of the free energy $E_{free}(t)$ of
AR 11158 in Fig.~9, along with the light curves of the GOES 1-8 \ang\
flux, which contains 36 flares above the GOES C0-class level during
the 2011 Feb 12-17 period. Note that the free energy calculated
with the COR-NLFFF code exhibits much stronger fluctuations than the
PHOT-NLFFF (Wiegelmann) code in the later days, while both codes
show an absence of significant fluctuations during the first day
(2011 Feb 12). 

We investigate now changes in the magnetic
energy $\Delta E_{free}(t)$ specifically during these 36 flare time
intervals. A sample of 9 flare time intervals are shown enlarged
in Fig.~10. We determined the flare start time ($t_{start}$),
peak time ($t_{peak}$), and end time ($t_{end}$) of these 36 events 
(Table 2) based on standard NOAA flare catalog, where the end time
is defined when the GOES flux drops to 50\% of the peak flux. 
We define a decrease of the free energy as follows, 
\begin{equation}
	\Delta E_{free} = (E_{post} - E_{pre}) =
	E_{free}(t=t_{post}) - E_{free}(t=t_{pre}) \ ,
\end{equation}
where the preflare reference time $t_{pre}$ is identified by the maximum
of the free energy during the preflare time interval [$t_{start}-0.3$ hr,
$t_{peak}$], and the postflare reference time $t_{post}$ is identified by 
the minimum of the free energy during the time interval 
$[t_{start}, t_{peak}+0.3$ hr]. The so-determined time intervals
[$t_{pre}, t_{post}$] of maximum energy dissipation during the risetime
of the flare (within a margin of 0.3 hr) are shown for both the
COR-NLFFF code (blue curves in Fig.~10) and the Wiegelmann NLFFF code
(red curves in Fig.~10). The uncertainties of the free energy,
$E_{free} \pm \sigma_E$, are empirically determined from the median
daily fluctuations of the free energy, which varied in the range of
$\sigma_E=(2 - 34) \times 10^{30}$ erg for the COR-NLFFF code, and 
$\sigma_E=(1.6 - 6) \times 10^{30}$ erg for the PHOT-NLFFF code.
The free energies and their uncertainties,
as well as the energy decrease ($\Delta E_{free}$) during the 36
flare events are listed in Table 2, for both the COR-NLFFF and the
(Wiegelmann) PHOT-NLFFF code.

The main result is that 29 out of the 36 flare events calculated
with our COR-NLFFF code exhibit a significant decrease in free energy
during the flare time interval, while only 7 events show no significant
change, either due to the smallness of the energy change or the large 
uncertainty of the method. Likewise we find a significant energy
decrease for most of the flares with the Wiegelmann NLFFF code,
but about a 3-10 times smaller amount of energy decreases. 
For instance, during the X2.2 flare 
on 2011-Feb-15, 02 UT (event \#16, Fig.~10 top right panel), we detect 
an energy decrease of $\Delta E_{free}=-(62 \pm 28) \times 10^{30}$ erg 
with the COR-NLFFF code, while about the half amount of 
$\Delta E_{free}=-(37 \pm 4) \times 10^{30}$ erg was detected
with the Wiegelmann NLFFF code (Sun et al.~2012a). 

In Fig.~11 we compare the decreases of free energy $(-\Delta E)$ with the
GOES 1-8 \ang\ fluxes and find a weak trend of a correlation between
the magnetic energy and the soft X-ray flux for the photospheric NLFFF method
(Sun et al.~2012), with a linear regression fit of $\log{(\Delta E)} 
\propto \log{(F_{GOES})}^{0.26\pm 0.06}$, while the coronal NLFFF method
shows no significant correlation. In flare models with
magnetic reconnection we would expect that the dissipated magnetic energy 
is correlated with the radiative output of heated plasma during flares.  
The scatterplot in Fig.~11 illustrates also that the energy decreases 
detected with the photospheric NLFFF code are about an order of magnitude
smaller than those with the coronal NLFFF code for small flares, and a
factor of three for the largest flares.

In Fig.~12 we show the correlation of potential energies (Fig.~12 left
panel) and nonpotential energies (Fig.~12 right panel) between the coronal NLFFF
code and the photospheric NLFFF code by Wiegelmann for 600 time steps (12-min cadence)
during the entire 5-day period. Both the potential and nonpotential energies correlate
well during the first 3 days (when the potential energy was $E_P \lapprox 10^{33}$
erg), but deviate during the last two days. At this time
it is not clear whether this discrepancy is due to some systematic center-to-limb
effect in the de-rotation of photospheric vector magnetograms (Sun et al.~2012a),
or due to a center-to-limb effect in the decomposition of LOS magnetograms (Section 3.1).
In principle, the coronal NLFFF code can represent vertically twisted field lines
at any longitude, if the magnetic charges are properly decomposed from the magnetogram.

A linear regression fit in Fig.~12 shows that the potential energy is recovered
by 85\% with the coronal NLFFF code, while the nonpotential energy is recovered
by 74\%. Since the magnetic energy scales with the square of the magnetic field
strength, the potential field is recovered by 92\%, using $n_{mag}=100$ magnetic
source components in our decomposition technique. This difference causes a
reduction of the free energy up to a factor of 4 for the coronal NLFFF code,
compared with the photospheric (Wiegelmann) NLFFF code (Fig.~8d), which could 
indicate a systematic effect of higher twist in the non-forcefree zones of the 
photosphere and lower chromosphere.

\subsection{	Loop Illumination Effects 	}

When we analyze the time evolution of the free energy $E_{free}(t)$ in the
time profiles shown in Figs.~9 and 10 (blue histograms), we see that almost none
of the flares exhibits the expected behavior of a near-constant high level
of free energy before the flare, which then decreases to a lower level during the
impulsive flare phase, so that a simple energy difference between the levels
before and after the flare could be used as a measure of the dissipated
magnetic energy. Instead we observe often a rapid increase of the free
energy at the beginning of the impulsive flare phase, which we interpret
as an illumination effect of highly twisted loop structures (sigmoids) 
that yield a high contribution of nonpotential energy to the free energy 
(time step $t_1 \mapsto t_2$ in Fig.~13).
In this scenario, it is still possible to measure the decrease of free energy
after the peak (time step $t_2 \mapsto t_3$ in Fig.~13), 
regardless whether the preflare level is low (for invisible
sigmoids) or high (after illumination, say when the flare-associated
chromospheric evaporation process fills the sigmoids with dense plasma.
Once a sigmoid gets brighter, it will be detected with our automated
loop detection scheme and will cause higher amounts of free energy in the
COR-NLFFF forward-fitting algorithm. 

In some cases we see multiple peaks of the free energy during the flare phase
(e.g., secondary peak in time step $t_3 \mapsto t_4$ in Fig.~13),
which are likely to be associated with spatially different twisted structures.
We can apply the same interpretation of an illumination effect during intervals of 
increasing free energy and magnetic dissipation during the time intervals of
decreasing free energy (time step $t_4 \mapsto t_5$ in Fig.~13). 
In order to caclulate the total dissipated energy, we
would have to add each decrease of the free energy during the impulsive flare
phase. If multiple such episodes overlap in time, even the sum of all free
energy decreases will be only a lower limit to the total dissipated energy,
because the time overlap will partially cancel energy increases (due to
illumination effects) and energy decreases. 

In summary, we conclude that
our method of estimating the dissipated free energy from the difference
between the peak of the free energy during the preflare phase and the minimum
after the impulsive flare phase will generally yield a lower limit.
This scenario may be tested in future work by measuring the illumination effects
and dissipation of free energy in spatially separated regions and with higher
time resolution, and this way may yield more accurate estimates of the total 
dissipated magnetic energy in flares. 

\section{		DISCUSSION					}

\subsection{		Previous Studies				}

The magnetic field of AR 11158 during a few days around the X2.2
flare on 2011-Feb-15, 02 UT, has been the subject of at least 40
different publications for the following reasons: (1) The X2.2 flare
was the first X-class flare observed with SDO; (2) the active region
has been observed near the central meridian, which is favorable for
any magnetic field extrapolation method, and (3) the HMI/SDO data have 
been calibrated and released (Hoeksema et al.~2014). 

\subsubsection{Previous Studies on the X2.2 flare}

The over 40 publications of this active region and its main X2.2 flare cover: 
the magnetic evolution as computed with nonlinear force-free codes 
 (Sun et al.~2012a; Jiang and Feng 2013; Inoue et al.~2013; Tarr et al.~2013)
and with stereoscopic comparisons (Wang et al.~2014),
MHD magnetic flux-rope modeling
 (Schrijver et al.~2011), 
the photospheric magnetic response to the flare
 (Wang et al.~2012, 2013; Liu et al.~2012, 2013; Petrie 2012a,b),  
magnetic field collapse and quasi-periodic loop oscillations during the flare
 (Gosain 2012; Dolla et al.~2012),
coronal waves
 (Schrijver et al.~2011; Olmedo et al.~2012), 
coronal mass ejection and interplanetary kinematics
 (Schrijver et al.~2011; Maricic et al.~2013a, 2013b),
velocity and magnetic transients of the flare
 (Maurya et al.~2012; Vemareddy et al.~2012a,b; Petrie 2012a,b), 
horizontal flow motion during the flare
 (Beauregard et al.~2012; Liu et al.~2013), 
the sunquakes and helioseismic response to the flare
 (Kosovichev 2011; Zharkov et al.~2011, 2013; Alvarado-Gomez et al.~2012),
rapid sunspot rotation during the flare
 (Jiang et al.~2012; Vemareddy et al.~2012a,b),
the EUV continuum evolution during the flare
 (Milligan et al.~2012), 
high-speed outflows shortly before the flare
 (Su et al.~2012),
flare-associated radio bursts 
 (Borovik et al.~2012; Chen et al.~2013; Yashiro et al.~2014),
magnetic modeling and flare trigger of a preceding M-class flare 
 (Kusano et al.~2012; Toriumi et al.~2013),  
and spectroscopy of flows in the kernel of a following M-class flare
 (Young et al.~2013). 

\subsubsection{	Previous Studies on Active Region NOAA 11158	}

Studies on AR 11158, which hosted the X2.2 flare, include 
the evolution of the nonpotentiality and helicity of the active region
 (Liu and Schuck 2012; Jing et al.~2012; Song et al.~2013; Inoue et al.~2013; 
  Tziotziou et al.~2013; Vemareddy 2012a,b), 
the quadrupolar magnetic configuration of the active region
 (Sun et al.~2012b),
the magnetic energy distribution of the active region
 (Shen et al.~2013), 
the magnetic calibration of photospheric Doppler velocities 
in the active region (Welsch et al.~2013), 
automated loop tracing, and emission measure and temperature analysis 
of the active region (Aschwanden et al.~2013a).

\subsection{	Energy Budget of the X2.2 Flare on 2011-Feb-15		}

A primary goal of this study is to establish the amount of energy that is
dissipated during solar flares, which we hope to derive from the change
of free magnetic energy during the flare time intervals. The largest flare 
during the analyzed time interval, the X2.2 flare on 2011 February 15, 02 UT, 
is the most suitable event in this context, since we have free energy 
calculations from multiple NLFFF codes: the Wiegelmann code (Sun et al.~2012a),
an MHD relaxation code (Jiang and Feng 2013), and our COR-NLFFF code based 
on forward-fitting to coronal loops.  In Table 3 we compile the different 
forms of energies that have been calculated for this flare.

The total (nonpotential) magnetic energy averaged from the four values 
listed in Table 3 has a mean of $E_{N} = (10.4\pm1.4) \times 10^{32}$ erg,
which corresponds to a mutual agreement within $\approx 15\%$. The potential
energies have a mean of $E_{P} = (8.7\pm0.8) \times 10^{32}$ erg and
agree within $\approx 10\%$. The free magnetic energy has a somewhat
larger scatter, within a range of $E_{free}=(1.65-2.4) \times 10^{32}$ erg
for the standard NLFFF codes, with or without pre-processing, while we find
with the COR-NLFFF code a similar value, i.e., 
$E_{free}=(0.98\pm0.20) \times 10^{32}$ erg.
However, the largest discrepancy is found for the decrease of free energy 
during the X2.2-class flare, for which a relatively low value of
$\Delta E=E_{post}-E_{pre}=(0.37\pm0.04) \times 10^{32}$ erg has been found 
with the Wiegelmann NLFFF code (Sun et al.~2012a), a mere $\approx 15\%$ 
of the available free energy, while we find about a double value for the
energy decrease of 
$\Delta E=E_{post}-E_{pre}=(0.62 \pm 0.28) \times 10^{32}$ erg
with our coronal forward-fitting code, which makes up about 60\% of the
free available energy. This is a very intriguing result, because it
sets a lower limit on the energy budget of the flare and may 
invalidate the results of some NLFFF codes.

We have some other complementary information on the energy input of this 
flare, from minimum-current corona modeling and from the virial theorem.
Using a magnetic charge topology (similar to the magnetic
source decomposition in our COR-NLFFF code, see Sections 2.2 and 3.1),
together with the minimum current corona model, the amount of reconnecting 
flux was calculated and a drop in the free magnetic energy of $\Delta E_{MCC}
=-1.68\times 10^{32}$ erg was found during the X2.2 flare (Tarr et al.~2013), 
which agrees within a factor of about two with our COR-NLFFF code
and the Wiegelmann NLFFF code (Sun et al.~2012a). The nonpotentiality can also
be calculated with the virial theorem (Chandrasekhar 1961; Low 1982),
which yields an energy drop of $\Delta E_{virial}=(1.05 \pm 1.04)
\times 10^{32}$ erg (Vemareddy et al.~2012b), and a similar amount of 
$\Delta E_{virial}=0.84 \times 10^{32}$ erg from an independent other
calculation (Tziotziou et al.~2013). Both of these values obtained with
the virial theorem are comparable with our forward-fitting method, but
exceed the value obtained with the Wiegelmann code (using pre-processed
data) by a factor of $\approx 3$ (Sun et al.~2012a). 

Another test of the plausibility of a free energy calculation is the
positive balance after energy losses during a flare. The major part of the
dissipated energy in flares is believed to be converted into acceleration
of nonthermal particles and heating of the chromosphere by precipitating
particles (ions and electrons). A calculation of the nonthermal energy
spectrum in the range of $E=25-50$ keV using RHESSI data yielded a 
value of $E_{HXR}=0.54 \times 10^{32}$ erg (Sun et al.~2012a), which 
exceeds the energy budget caclulated with the Wiegelmann code
($\Delta E = (0.37 \pm 0.04) \times 10^{32}$ erg), while it dissipates less 
than the magnetic energy budget calculated with our forward-fitting code
($\Delta E = (0.62 \pm 0.28) \times 10^{32}$ erg).

On the other hand, the total thermal energy produced in a flare can also
be estimated from the differential emission measure (DEM) distribution
obtained in soft and EUV wavelengths, for which we find at the flare
peak, using AIA/SDO data, a total thermal energy of $E_{th}=0.2 \times 10^{32}$
erg (which makes about 30\% of the energy budget), with a peak temperature 
of $T_p=17.8$ MK, a peak electron density of
$n_p=3.2 \times 10^{11}$ cm$^{-3}$ (assuming a filling factor of unity), 
and a flare radius of $L_p=16.3$ Mm
(Aschwanden et al.~2013a). Other forms of energy losses have been
determined in Ly$\alpha$, soft X-rays, Ly continuum, etc. (Table 3)
with EVE/SDO (Milligan et al.~2012), which are all smaller than 
$\approx 1\%$ of the free magnetic energy. 

In summary, the decrease in free energy during the X2.2 flare is 
consistently determined to be $\Delta E \approx (0.6-1.0) \times 10^{30}$ erg
with our forward-fitting code and with the virial theorem. 
The estimated energy losses due to acceleration of nonthermal particles 
and heating of the thermal plasma do not exceed this energy budget. 
The same conclusion holds for the 5 M-class flares that occurred in the
same active region during the time period of 2011 Febr 12 to 17, for which
we have energy changes calculated with the virial theorem
(Vemareddy et al.~2012b; Tziotziou et al.~2013), and for which we have
calculations of the total thermal energy (Aschwanden et al.~2013a), as
compiled in Table 4. For additional comparisons of the energy budget in
flares see also Fig.~4 in Emslie et al.~(2012), and Emslie et al.~(2004, 2013).
In the latter studies, however, the free magnetic energy is not determined,
but only the total potential field energy is given instead, which is about
an order of magnitude larger than the free energy that is available for
energy dissipation during flares. Therefore, our study provides much 
stronger contraints on the upper limit of the flare energy budget.

\subsection{	Photospheric versus Coronal Magnetic Field Constraints	}

This is the first study that quantitatively compares the coronal magnetic
field computed by photospheric (extrapolation) and by coronal (forward-fitting)
methods. In principle we aim to calculate the same nonlinear force-free
field solution in a given computation box, but the significant differences 
we find here may provide important information on systematic errors of each 
magnetic field calculation method, or on the non-forcefreeness in the lower 
chromosphere. A recent review on the success and future improvements of
NLFFF codes is given in Regnier (2013).

\subsubsection{		Comparison and Discrepancies 		}

Here we discuss a quantitative comparison between the photospheric 
NLFFF extrapolation code (Wiegelmann (2004), as applied in Sun et al.~(2012a),
and the coronal forward-fitting (COR-NLFFF) code used in this study. In Fig.~8 
we show the evolution of the free energy over a time interval of 5 days 
computed in 1200 time steps (Fig.~8d), along with the potential 
(Fig.~8b) and nonpotential energy energy 
(Fig.~8c). We show also the correlation of the potential $E_P$ 
and nonpotential energy $E_N$ between the two codes in Fig.~12.
We make the following findings: (1) The potential magnetic energy agrees
with good accuracy between the two codes, at least during the first three 
days when the active region is near disk center. It is not clear whether 
the deviations during the last two days is due to a center-to-limb effect
or due to a difference in the computation box. (2) The free energy obtained 
with the Wiegelmann NLFFF code exceeds that of the COR-NLFFF code up to a 
factor of 4 (Fig.~8d); (3) The decrease of the free energy measured with
the Wiegelmann NLFFF code during flares is reduced by an amount of up to an 
order of magnitude, compared with the COR-NLFFF code. The energy descrease 
falls short of the nonthermal energy required to accelerate hard X-ray 
producing electrons during the X2.2 flare. On the other side, the 
free energy decrease during the X2.2 flare obtained with the COR-NLFFF code, 
$\Delta E_{free} \approx 0.6 \times 10^{32}$ erg, 
is consistent with the values obtained with the virial theorem (Vemareddy 
et al.~2012; Tziotziou et al.~2013). (4) The time profile of the free energy 
obtained with the Wiegelmann code exhibits about an order of magnitude less 
variability than with the COR-NLFFF code. In the following we discuss different 
effects that possibly could explain these discrepancies between the two codes.

\subsubsection{		Spatial Resolution			}

Both codes are using the original HMI data with a pixel size of
$0.5\arcsec$. The HMI data used for the Wiegelmann NLFFF code are averaged
down to two pixels ($1.0\arcsec$) (Sun et al.~2012a), while the data used for 
the COR-NLFFF code are rebinned to three HMI pixels ($1.5 \arcsec$), for which 
the optimum match in the Gaussian decomposition of the line-of-sight
magnetograms was established (Fig.~15a). It was speculated (Sun et al.~2012a) 
whether small-scale fields with large gradients could cause electric currents 
that are unresolved with the averaged HMI data, which could lower the 
resulting free energy. However, the spatial resolution of HMI data should 
affect both codes in a similar manner, since both codes used averaging of
two or three pixels, and thus the averaging cannot explain the discrepancy
of the obtained free energy between the two codes. 

\subsubsection{		Computation Box				}

One constraint of standard NLFFF codes is the planar computation box
with cartesian geometry, which requires a remapping of an active region at
heliographic position $(l,b)$ to disk center $(0,0)$. The remapping
transforms a fraction of the transverse field to the line-of-sight field
component, which scales with the sine-function of the center-to-limb distance.
In the study of Sun et al.~(2012a), re-mapping of the HMI magnetogram
using the Lambert equal area projection has been used, which is different
from the COR-NLFFF code, where the full sphericity of the Sun is taken into
account and no remapping is needed. However, since the X2.2 flare occurred 
near disk center, this effect should be very small and cannot explain the 
discrepancy in the free energy.

\subsubsection{		Time resolution 			}

For the forward-fitting to coronal loops we used the HMI 45 s data,
which were processed by the HMI team from 135 s time intervals, so
we see variability of the magnetic field down to about 2 minutes.
The time step in modeling of AIA data is 6 minutes, but the exposure
times of AIA are typically $\approx 2$ s. We ran the Wiegelmann NLFFF
code with a time step of 12 minutes, which was processed by the HMI
team from 1350 s (22.5 minute) time intervals. Thus part of the much
lower modulation depth of the free energy calculated with the 
Wiegelmann NLFFF code could be due to the poorer time resolution.
The data noise in the HMI data with 1350 s would be, compared with
the HMI 135 s data used for coronal forward-fitting, a factor of
$\sqrt{10} \approx 3.3$ lower. However, since we detect about a
factor of 10 less decrease of free energy during flares between
the two methods (Fig.~11), there is still a factor of 3 less
modulation depth unexplained.

\subsubsection{		Pre-Processing of HMI Data		}

Before using the Wiegelmann NLFFF code, the photospheric vector magnetograph 
data were ``pre-processed'', a procedure that minimizes the flux, force, 
and torque of the 3D magnetic field vectors (Wiegelmann et al.~2006). 
It is suspected that the pro-processing may introduce too much smoothing 
(or time averaging) in the photospheric data, which could reduce the 
electric currents and this way causes an underestimate of the free energy 
(Sun et al.~2012a). 
In the study of Jiang and Feng (2013), the pre-processed data yield about 
$10\%$ smaller total and potential energies than the raw data, resulting 
into about $15\%$ less free energy for the X2.2 flare. This small
difference cannot explain why we measure a factor of 2 different
decrease in the free energy during the X2.2 flare. 

\subsubsection{		Coronal Loop Geometry			}

The COR-NLFFF code performs forward-fitting of an approximate NLFFF solution 
to the shapes of coronal loops. This code is designed to fit the observed loop 
geometries, and thus retrieves the helical twist of loops. 
The helical twist of magnetic field lines plays a key role before
large flares, such as observed here, which was confirmed with a 3D NLFFF
magnetohydrodynamic (MHD) relaxation method (Inoue et al.~2013). The
latter method determined one-half to one-full twist before the M6.6 and
the X2.2 flare (Inoue et al.~2013), which clearly untwisted after
the X2.2 flare, as it can be seen in the  northern portion of AR 11158 
in Fig.~7, depicted in two forward-fits before and after the flare peak.

The accuracy of the best-fit solutions of the COR-NLFFF code is typically
a median misalignment angle of $\mu_2 \approx 4.5^\circ$ (Fig.~8 bottom),
which corresponds to a 3D misalignment angle of 
$\mu_3=\mu_2 \sqrt{(3/2)} \approx 5.5^\circ$. 
In contrast, standard NLFFF codes, which by definition do not use any 
information of the observed coronal geometry, end up with typical 
3D misalignment angles in the range of $\mu_3 \approx 24^\circ-40^\circ$ 
(DeRosa et al.~2009).
The misalignment angle $\mu_3$ is directly related to the force-free
parameter $\alpha$, and thus to the free energy. If a misalignment angle 
$\mu$ increases to the double value in a bad fit, this translates into a 
doubled azimuthal (nonpotential) magnetic field component $B_\varphi \propto 
B_{pot}\tan{(\mu)}$, an thus to a free energy ratio squared, i.e., 
$q_{free}=(B_{\varphi}/B_{pot})^2 = \tan{(\mu)^2}$. For instance, if a NLFFF 
solution is misaligned by $\mu = 30^\circ$, the free energy increases 
from $E_{free}/E_{pot}=\sin{(30^\circ)}^2 \approx 0.25$ to a value of
$E_{free}/E_{pot}=\sin{(60^\circ)}^2 \approx 0.75$, or the total
nonpotential energy changes from $E_{N}/E_P=1.25$ to $E_{N}/E_P=1.75$,
which represents an increase of $40\%$. This example may explain part of the
discrepancy in free energies, which are about a factor of 4 higher
when determined with a photospheric extrapolation code (such as the 
Wiegelmann NLFFF code used in Sun et al.~2012a), compared with the
COR-NLFFF code that performs forward-fitting to coronal loops. 
The most plausible solution for the observed discrepancy in misalignment
angles is that the photospheric magnetic field has about a factor of 
$\sqrt{4} = 2$ higher azimuthally twisted field components in 
the non-forcefree zones than the coronal loops in the forcefree corona, 
which would produce a factor of 4 higher free energy for extrapolation 
codes that use the photospheric transverse field (which mostly contains
the azimuthal twisted field component). Consequently, the coronal loops
do not have a uniform twist all the way down to the photosphere, but
increase their twist angle by a factor of about 2 at the photospheric
footpoints. 

\subsubsection{		Non-forcefree Flare Dynamics 	}

Another possible reason for the apparent underestimate of the free energy
during the X2.2 flare using the Wiegelmann code has been attributed
to the dynamics of the eruption, which could possibly produce a strong 
deviation from a force-free state during the postflare phase (Sun et al.~2012a).
However, the non-forcefreeness during the eruption should affect both the
Wiegelmann NLFFF code and the COR-NLFFF forward-fitting method in a similar
way, because both codes calculate the NLFFF solution under the assumption
of force-freeness, and thus it cannot explain the discrepancy in the free
energy between the two codes. Moreover, the free energy calculated from
photospheric data is up to a factor of 4 larger than the values
calculated from forward-fitting of coronal data (Fig.~8d), measured
at 1200 time steps during the entire time period of 5 days, and thus
the discrepancy in free energies is persistently present, not just only
during times of flaring or CMEs. However, inspecting Fig.~12b, there
is a saturation apparent when the photospheric nonpotential energy reaches
the highest values of $E_N \approx 1 \times 10^{33}$ erg, while
the coronal free energies can exceed this limit, which may indicate
a photospheric line-tying effect.

\subsubsection{	  Temporal Variability and Helicity Flux  	}

A striking difference is the time evolution of the free energy, as calculated
at 1200 time steps over 5 days, is the degree of flucutations in the
time profile $E_{free}(t)$ between the two codes (Fig.~8c, 8d). The time
profile of the coronal COR-NLFFF code shows about an order of magnitude larger 
modulation depth than the photospheric Wiegelmann NLFFF code. 
Are those fluctuations real? The largest peaks and the decreases of the
free energy $E_{free}(t)$ do indeed correlate with flaring events, such 
as evidenced by the GOES light curves (Figs.~8a) and the expanded
time profiles of $E_{free}(t)$ (Fig.~10). 

We compare the evolution of the nonpotentiality and helicity 
in the active region 11158 (from Fig.~12 and 14 in Liu and Schuck 2012).  
We show a juxtaposition of the time profiles of the free energy 
(obtained with the coronal COR-NLFFF code) and the helicity flux
(Liu and Schuck 2012) in Fig.~14, which agree in the general trend
of a systmatic increase over the 5 days of observations, while
detailed correlations cannot be established due to the noise in the
helicity flux (estimated to be 23\% by Monte-Carlo simulations with 
the DAVE4VM code of Schuck 2008). In contrast, the photospheric
(Wiegelmann) NLFFF code (Fig.~14a) shows a peak of the free energy 
at the beginning
of the third day (during the largest flare), and then decreases during
the next two days, which is different from the evolution of the coronal
free energy (Fig.~14b) and the helicity flux (Fig.~14c).

\section{		CONCLUSIONS					}

In this study we improved the accuracy and performance of the 
{\sl Coronal Non-Linear Force-Free Field (COR-NLFFF)} forward-fitting 
code substantially by (1) implementing an automated loop 
tracing code for detection of coronal loops in multi-wavelength
EUV images, which makes the manual or visual loop tracing unnecessary 
(Aschwanden 2010), and (2) by optimization of the forward-fitting
technique to 2D loop coordinates, which relinquishes 3D reconstruction
with stereoscopy (Aschwanden 2013c). In this study we applied the
COR-NLFFF code to magnetic field modeling of AR 11158 during the time
interval of 2011 Feb 12 to 17, which includes an X2.2 flare plus 
35 M and C-class flares. We calcuated the free magnetic energy
over the 5 days with a cadence of 6 minutes and compare the results
with standard NLFFF calculations using the Wiegelmann code with a
cadence of 12 minutes. We compare quantitatively the magnitude and
evolution of the free energy during the GOES-detected flares, in
particular the detected decreases of free energy before and after
the flares with both types of codes. The standard NLFFF codes
(Wiegelmann 2004; Jiang and Feng 2013) use extrapolation of 
photospheric vector magnetograph data (with or without pre-processing),
while our forward-fitting COR-NLFFF code uses coronal
loop constraints. We compare also calculations with the virial theorem. 
The results are the following:

\begin{enumerate}
\item{The total nonpotential magnetic energy measured during the
X2.2 flare on 2011 Febr 15, 02 UT, agrees well ($\approx 15\%$) 
between photospheric standard NLFFF codes and the coronal 
forward-fitting code, i.e., $E_{N} = (10.4\pm1.4) \times 10^{32}$ erg.
However, the total energy determinied with the virial theorem is 
about a factor of two lower, i.e., $E_{N}=(4.97\pm1.58) \times 10^{32}$ erg, 
probably due to a different computation box.}

\item{The potential energy measured during the X2.2 flare agrees
also well ($\approx 10\%$) between the two types of codes, i.e.,
$E_P = (8.7\pm0.8) \times 10^{32}$ erg. The small difference, which is 
commensurable with the calibration of HMI and uncertainties among
different NLFFF codes, corroborates the accuracy of the potential field 
modeling approach in terms of a a limited number of decomposed magnetic 
souces. The potential energy calculated with the virial theorem, 
$E_{P} = (2.43\pm0.78) \times 10^{32}$ erg, is a factor of
3 smaller, similar to the nonpotential energy, probably due to a 
different computation box.}

\item{The free magnetic energy measured during the X2.2 flare, i.e.,
the difference between the nonpotential and potential energy, disagrees 
by a factor of about two between the two types of codes. 
For standard NLFFF codes we find a value of in the range of 
$E_{free}=(1.65-2.4) \times 10^{32}$ erg, varying by about $15\%$ 
with or without pre-processing, while we find a similar value of 
$E_{free}=1.0 \times 10^{32}$ erg with the coronal forward-fitting 
COR-NLFFF code, so there is almost agreement between all codes in this case.}

\item{The most critical quantity is the decrease of free magnetic energy
before and after the X2.2 flare, which sets an upper limit on the energy
budget of flares. We find a decrease in the free energy by 
$\Delta E=E_{post}-E_{pre}=(0.37\pm0.04) \times 10^{32}$ erg with 
the standard photospheric NLFFF code (Sun et al.~2012a), which is about half
of the decrease measured with the coronal forward-fitting COR-NLFFF
code, i.e., $\Delta E=E_{post}-E_{pre}=(0.62\pm0.28) \times 10^{32}$ erg, 
which is also consistent with the energy drop calculated with the virial 
theorem i.e., $\Delta E=E_{post}-E_{pre}=(1.05\pm1.04) \times 10^{32}$ erg
(Vemareddy et al.~2012), or $\Delta E=E_{post}-E_{pre}=0.84 \times 
10^{32}$ erg (Tziotziou et al.~2013). The fact that photospheric NLFFF
codes measure a higher amount of free energy than the coronal NLFFF codes,
but a too low decrease during flares, could partly be due to a smoothing 
and time-averaging effect of the pre-preprocessing algorithm.}

\item{The time evolution of the free energy, as calculated with a
time resolution of 12 minutes over 5 days with both types, exhibits
a systematic discrepancy in the amount of free energy obtained with
the photospheric NLFFF versus the coronal NLFFF code. Since the free energy 
is dominated by the nonpotential transverse photospheric field, which
corresponds to the azimuthal field component $B_{\varphi}$ in helically 
twisted field lines, it is conceivable that the twist of field lines is not
uniform along their length, but could be stronger twisted by about a factor of
$\sqrt{2}$ in the non-forcefree photosphere, and this way could explain
why standard NLFFF codes produce a factor of $\approx 2$ higher free
energies than coronal-fitting NLFFF codes.}

\item{Among the 36 GOES C,M, and X-class flares, we find a significant
decrease in the free magnetic energy during the flares in 29 cases
with the COR-NLFFF code. 
Similarly we find a significant decrease in many cases with the
photospheric NLFFF code, however the energy drop is about a factor
of $\approx 10$ smaller for the photospheric NLFFF code than for the coronal
(COR-NLFFF) code, and thus confirms statistically the same trend 
as we found for the X2.2 flare.}

\item{The time evolution of the free energy in time steps of 12 minutes
over 5 days exhibits about an order of magnitude stronger modulation
depth for the coronal NLFFF code than for the photospheric NLFFF code,
which is likely to be produced by the same effect that produces
smaller energy decreases during the 36 flares, which could be partially
be attributed to the smoothing effects of the pre-processing algorithm.} 

\end{enumerate}

This study represents the first comparative test case between
photospheric and coronal NLFFF codes, which revealed discrepancies
resulting from the inconsistency between the photospheric non-forcefree
and the coronal force-free field. It appears that the so far existing
preprocessing algorthm that supposedly optimizes the photospheric
field into a force-free field, is not consistent with the coronal
force-field extrapolated downward to a photospheric altitude level.
Therefore we conclude that the coronal field cannot be retrieved from
photospheric information alone, but requires explicit information from
the magnetic field in the force-free corona. While the present 
COR-NLFFF code used here is based on an analytical approximation that is
accurate two second order, more advanced NLFFF codes that include
the geometry of coronal loops should be developed in future.
One code in this direction, using a quasi-Grad-Rubin scheme is
under current development (Anna Malanushenko, private communication),
using visually traced loops as input. Of course, ultimate objectivity
can only be achieved with automated loop recognition algorithms. 


\section*{Appendix A: Parametric Study of the Numeric COR-NLFFF Code }

The default control parameters of the {\sl Coronal Non-linear Force-Free Field
(COR-NLFFF)} forward-fitting code used here are listed in Table 1. 
The accuracy of the forward-fitting results is primarily quantified 
with the median 2D misalignmen angle $\mu_2$ between the theoretical 
and observed (loop-aligned) magnetic field, obtained from the best 
forward-fit (which should converge to small values in the optimum case). 
Another test quantity is the ratio of the nonpotential to the potential
energy, $q_E=E_{N}/E_P$, which should be a slowly-varying function 
of time during non-flaring episodes, and is expected to display a decrease
during flares, the time interval when magnetic energy is dissipated. 
Therefore, in the following parametric study we demonstrate how 
these two quantities $\mu_2$ and $q_E$ vary as a function of the numerical
input parameterers (Figs.~15 and 16). 

All tests of this parametric study are performed with AIA and HMI data from 
2011-February-15 in the time interval starting at 00:00:00 UT and 
ending 12 s later, in a field-of-view of $x=[-0.0268, 0.3268]$ solar radii 
in East-West direction and $y=[-0.3885,-0.0885]$ in South-North direction.
Since the COR-NLFFF code takes the full sphericity of the Sun into account, 
no transformation to Sun center is needed, as it would be required for most 
traditional NLFFF codes with a planar boundary of the computation box
centered at Sun center.

\underbar{\sl Spatial resolution of magnetogram $\Delta x_{mag}$ 
(Figs.~15ab, 16ab):} What is the ideal HMI resolution for our code?
The HMI magnetogram has a full resolution of $\Delta 
x=0.00052$ solar radii. We rebin the magnetogram by factors of 1 to 10 
HMI pixels and show the obtained non-potential energy $E_{N}$
and potential energy $E_P$ as a function of the rebinned spatial resolution 
$\Delta x_{mag}$ (Fig.~15a) and their ratio $q_E=E_{N}/E_P$ (Fig.~15b),
respectively.
We see a sharply peaked function $E_{N}(\Delta x_{mag})$ that peaks at
$\Delta x_{mag}=0.0015$ (or 3 HMI pixels). We can understand this
function as a combination of two effects, i.e., (i) the undersampling of 
overresolved structures in the magnetogram in the case of $\lapprox 3$ 
HMI pixels, and (ii) over-smoothing of unresolved structures in the case of 
$\gapprox 3$ HMI pixels. The undersampling results because of the limited
number of modeled magnetic source structures 
(typically $n_{mag} \approx 100$) in
magnetograms, which cover a full-resolution size of $577 \times 577 = 332,929$ 
pixels, with a sizable fraction $n_{pixel}$ containing a significant magnetic 
field strength. The limitation of $n_{mag} \ll n_{pixel}$ of the code
causes an underestimate of both the potential as well as the 
nonpotential magnetic energy at high spatial resolution. 
If we approximate the area with high 
magnetic field strengths with a mean value $B$, the total magnetic energy is
expected to scale as $E_{mag} = n_{mag} B^2 \Delta x_{mag}^2$ for
$n_{mag} \ll n_{pixel}$, which is a quadratic function of the pixel size
and fits the observed function in the range of $\Delta x_{mag}=1,...,3$
pixels (left red curve in Fig.~15a). On the other side, above the critical
limit where the number of model components is sufficient to cover the 
number of macropixels, i.e., $n_{mag} > n_{pixel}$, we expect that the
rebinning conserves the magnetic flux $\Phi = \sum B \Delta x_{mag}^2$,
but the rebinned field strength $B_{rebin}$ scales then reciprocally
with the macropixel size, i.e., $B_{rebin} = B / \Delta x_{mag}$, which
leads to a quadratic decrease of the magnetic energy as a function of
the (rebinned) macro pixel size, i.e., $E_{mag} \propto (\Delta x_{mag})^{-2}$
(right red curve in the range of 3-10 HMI pixels in Fig.~15a). 
Therefore, the least biased
magnetic energy is measured at the critical limit $n_{mag} \approx n_{pixel}$,
which turns out to be 3 HMI pixels here (or $\approx 1.5\arcsec \approx
1500$ km). In other words, the critical scale of $\Delta x_{mag} \approx 
1500$ km represents the optimum scale where the number of model sources match
the number of significant magnetic sources, so that the side effects of
under-resolving and over-resolving dissappear. Hence we are using
rebinned magnetograms with $\Delta x_{mag}=3$ HMI pixels throughout
this study.

The coverage of the fitted magnetic
areas can be seen from the circles displayed in the bottom panels of Fig.~17,
for three different resolution scales (1, 3, and 10 HMI pixels).
The spatial resolutions shown in Fig.~17 are ($\Delta x_{mag}=0.0005, 0.0015$,
and 0.005 solar radii). We find a maximum in the detection of nonpotential 
magnetic energy of $E_{N}=8.6 \times 10^{32}$ erg,
a potential energy of $E_P=7.6 \times 10^{32}$ erg,
a free energy of $E_{free}=E_{N}-E_P= 1.0 \times 10^{32}$ erg,
which corresponds to an energy ratio of $q_E=1.13$ (Fig.~15b), 
or a free energy of 13\% of the potential energy.
These values agree with other studies using the traditional NLFFF 
method, i.e., $E_{N} \approx 10 \times 10^{32}$ erg (Sun et al.~2012a;
Jiang and Feng 2013). The mean misalignment angle of the best-fit 
non-potential field using our COR-NLFFF model is $\mu \approx 5.0^\circ$ 
(Fig.~16b), while the potential field has a misalignment of
$\mu \approx 10.0^\circ$ (Fig.~16a).  

\underbar{\sl Number of magnetic sources $n_{mag}$ (Fig.~15c, 16c):}
What is the optimim number of magnetic sources that is needed for the 
decomposition of the LOS magnetogram and construction of the model map?
In the three examples shown in Fig.~17 we used $n_{mag}=100$ magnetic 
sources, which appears to be sufficient to represent most of the magnetic
flux down to a level of a few percent, even for this case of a complex
active region two hours before an X-class flare. We varied the number
of decomposed magnetic sources from $n_{mag}=10$ to $n_{mag}=200$
and found a quite robust behavior in the best-fit value of the
nonpotential energy ratio $q_E \approx 1.13$ for $n_{mag} \gapprox 50$
(Fig.~15c), with a mean misalignment angle of $\mu \approx 5.0^\circ$ 
(Fig.~16c). We choose $n_{mag}=100$ as default value.

\underbar{\sl Threshold level in loop detection $q_{med}$ (Fig.~15d, 16d):} 
The threshold level of automated loop detection in different wavelength
filters is empirically determined by visual inspection of false
loop detections in noisy image areas. We find a good discrimination
for the following factors of the median flux (obtained from the boundary 
areas of
the highpass-filtered EUV images): $q_{med,0}(\lambda)=[5, 2, 0, 0, 1, 5, 5]$
for the wavelengths $\lambda=[94, 131, 171, 193, 211, 304, 335]$ \ang .
In our parametric study we vary these median levels by a factor of
$q_{med}=0.2, ...,2.0 \times q_{med,0}$. The obtained energy ratios 
$q_{free}$ (Fig.~15d) and misalignment angles $\mu_2$ (Fig.~16d) show
almost invariant values as a function of the flux threshold variation,
which corroborates the robustness of our loop detection scheme. 

\underbar{\sl Loop selection by wavelength $\lambda$ (Fig.~15e, 16e):} 
We vary now the parameters that affect the selection of loops (Section 3.2). 
Let us first consider the different wavelength filters ($\lambda$).
The nonpotential energy ratios $q_E=E_{N}/E_P$
retrieved from each single wavelength separatetly are shown in Fig.~15e,
which obviously demonstrates that not each wavelength contains a 
representative subset of nonpotential loops. If we combine the loop
tracings from all 7 wavelength filters, we find a
value of $q_E = 1.13$ (or 13\% free energy), which is approximately 
retrieved by those wavelength filters  
that contain representative subsets of the loops (Fig.~15e). 
However, some filters yield values that deviate substantially from
the mean, and thus seem not to contain a representative sample of loops.
Investigating the misalignment angles $\mu_2$ for these seven filters 
(Fig.~16e), we see that the best-fit values in each wavelength converge
to a range of $\mu_2 \approx 4^\circ-7^\circ$, while the synthesized set 
of loops from all wavelengths yields $\mu_2 \approx 5^\circ$.
We conclude that a well-balanced sample of loops synthesized 
from all wavelength filters is required to obtain the correct free energy.

\underbar{\sl Maximum number of loops per filter $n_{loop}$ 
(Figs.~15f, 16f):}
If we combine all seven filters, we obtain about 1000 loop structures.
If we limit the maximum number of loops extracted per wavelength
filter to $n_{loop}=100$, we obtain $452$ loops, which reduces
the computation time by a factor of two. The parametric study
shown in Fig.~15f shows that we can reduce the number down to
$n_{loop} \approx 40$ without changing the result of the
nonpotential energy. Only below that limit we loose too much
potential loops (mostly in the 171 and 193 \ang\ images) so that
the sample is not representative anymore. Thus we choose 
$n_{loop}=50$ as default value. The misalignment angle is 
nearly insensitive to the maximum number of selected loops per
wavelength (Fig.~16f).

\underbar{\sl Minimum length of selected loops $l_{min}$ 
(Figs.~15g, 16g):}
Besides the minimum loop length criterion defined in
the loop tracing algorithm, we can additionally set a minimum
loop length criterion $l_{min}$ for the subset
of loops selected for forward-fitting. This allows us to balance
the relative ratio of short loops (which occur in the core of
active regions and exhibit generally a higher nonpotential energy)
and long loops (which occur in the outer shells of active regions
and exhibit a more potential field characteristics). The parametric
study in the range of $l_{min}=[10,100]$ pixels exhibits 
for $l_{min} \lapprox 20$ pixels an overabundance of short loops,
producing a relatively high nonpotential energy ratio $q_E \approx 1.2$,
and inversely, an overabundance of long loops for $l_{min} \gapprox 40$ 
pixels, which are mostly potential-like and produce a too low nonpotential 
energy ratio $q_E \lapprox 1.1$ (Fig.~15g). 
Thus, we choose an intermediate value of $l_{min}=30$ as default.
The misalignment angle varies in the range of $\mu_2 \approx
3^\circ - 6^\circ$ in this parameter range (Fig.~16g).

\underbar{\sl Loop curvature radius $r_{min}$
(Figs.~15h, 16h):} The automated loop tracing code picks curvi-linear 
structures with a minimum
curvature radius $r_{min}$, while structures with shorter curvature
radii often result from ``curved chains'' of moss structure 
(Berger et al.~1999).
Tests with curvature radii varied from $r_{min}=10$ to 60 pixels
reveal no systematic tendency of the best-fit non-potential energy
ratio $q_E=E_{N}/E_P$ (Fig.~16h) or misalignment angle .
Thus we choose $r_{min}=30$ pixels as default value. 

\underbar{\sl Suppression of false loop structures:}
The most common false loop structures are: (1) curved chains of
moss features, (2) saturated pixels during strong flares,
(3) vertical streaks from pixel bleeding in the CCD camera during
strong flares, and (4) diffraction patterns from the entrance mesh filter
during intense flares. Features (1) and (4) can mostly be filtered out
by a minimum loop length requirement (which is choosen here to be
$l_{min} \ge 30$ AIA pixels). Loop structures containing saturated 
pixels are rejected based on their flux value ($F > 2^{14}-1$ DN/s).
Vertical streaks from pixel bleeding are also easily detected by
their small variation in x-coordinates (less than one pixel, either
in the northern or southern half of the automatically detected
structures). The rejection of false loop structures improves the
best-fit solution significantly, as we tested in images during
flare peaks. 

\underbar{\sl The number of loop segment points $n_{seg}$ 
(Fig.~15i, 16i):}.
Forward-fitting is carried out at $n_{fit}=n_L \times n_{seg}$ loop
2D positions $(x_i, y_i)$, $i=1,...,n_{fit}$, where $n_f$ is the
number of selected loops, and $n_{seg}$ is the number of loop
segments per loop, distributed equidistantly along the automatically
traced loop segments. A too small number of loop segments may not be
sufficient for a representative fit, and a too large number consumes
more computation time. The parametric study in Fig.~15i and 16i shows that
the nonpotential energy is robustly retrieved for $n_{seg}=3$
to 15. We choose $n_{seg}=7$ as default value.

\underbar{\sl The force-free $\alpha$-parameter increment $\Delta \alpha$
(Fig.~15j):}
The $\alpha$-parameter is varied for each magnetic source within a range
of $\pm\Delta \alpha$ for every magnetic source in each iteration cycle.
If we choose $\Delta \alpha$ too small, we need a too large number of
iterations to converge, requiring too much computation time. If we choose it
too large, the minimization of the misalignment angle per iteration
misses the local maximum and will fluctuate erratically instead of
converging steadily to the absolute minimum. The parametric study in
Fig.~15j and 15j show that the nonpotential energy is stably retrieved in the
range of $\Delta \alpha \approx 5-100$ [$R_{\odot}^{-1}$].
We choose $\Delta \alpha = 10 [R_{\odot}^{-1}]$ as default value.

\underbar{\sl The altitude range of the computation box $h_{max}$ 
(Figs.~15k and 15k):}
The height of the computation box can possibly affect the forward-fitted
loop solutions, because larger heights contain more ambiguities in
the reconstruction of the line-of-sight coordinate $z_i$ for each of
the fitting points. In addition, increasing the altitude yields a 
slightly higher amount of volume-integrated magnetic energy. The
parametric study shown in Figs.~15k and 15k, however, shows an invariant
retrieval of the nonpotential energy for altitude ranges of
$h_{max}=0.05-0.25$ solar radii. We choose $h_{max}=0.15$ as default
value.

\underbar{\sl The maximum number of iterations $n_{iter}$ 
(Figs.~15h, 16h):}
Our chosen forward-fitting method is the Powell method, which
determines the local minimum for each free parameter $\alpha_j$
sequentially per iteration cycle. An example of the change of misalignment
angle $\mu$ and nontpotential energy ratio $q_E=E_{N}/E_P$ as a
function of the iteration cycle is shown in Fig.~15l and 15l, for
$n_{iter}=1-10$. The misalignment angle starts with $\mu\approx 9^\circ$
for the potential-field solution in the first iteration, and converges
to a final best-fit value of $\mu\approx 5^\circ$ after the fourth iteration.
Thus we choose $n_{iter}=6$ as default value, since more iterations
were found not to improve the goodness-of-fit significantly.

\acknowledgements
The author appreciates the constructive comments by an anonymous referee 
and helpful discussions with Allen Gary, Anna Malanushenko,
Marc DeRosa, and Karel Schrijver. Part of the work was supported by
NASA contract NNG 04EA00C of the SDO/AIA instrument and
the NASA STEREO mission under NRL contract N00173-02-C-2035.

\clearpage


\begin{deluxetable}{lll}
\tabletypesize{\normalsize}
\tablecaption{Default control parameters of the coronal non-linear
force-free field (COR-NLFFF) forward-fitting code used in this study.}
\tablewidth{0pt}
\tablehead{
\colhead{Task:}&
\colhead{Control parameter}&
\colhead{Value}}
\startdata
Data selection:	&Date and time of observation	&2011 Feb 15, 00:00:00 UT \\
		&Instruments			&AIA, HMI (SDO)		\\
		&East-west field-of-view	&[ 0.0268, 0.3268]$R_{\odot}$ \\
		&North-south field-of-view	&[-0.3885,-0.0885]$R_{\odot}$ \\
		&AIA wavelengths		& 94, 131, 171, 193, 211, 304, 335 \ang\\
		&spatial pixel size 		& $\Delta x_{AIA} = 0.6\arcsec$ \\
Loop tracing:	&Lowpass filter boxcar		& $n_{sm1}=5$ pixels \\		
		&Highpass filter boxcar		& $n_{sm2}=7$ pixels \\		
		&Image base level/median& 	$q_{med}=5,2,0,0,1,5,5$ \\
		&minimum loop length    	& $l_{min}=30$ pixels \\
		&minimum loop curvature radius 	& $r_{min}=30$ pixels \\
		&field line step		& $ds_{field}=0.002 R_{\odot}$ \\
Magnetic sources: &rebinned pixel size		& $\Delta x_{mag}=3$ pixel (1.5\arcsec, 0.0015 $R_{\odot}$) \\ 
		&number of magnetic sources	& $n_{mag}=100$	\\
Loop selection: &maximum number per filter	& $n_{loop}=50$ \\
Forward-Fitting:&minimization iteration method 	& Powell \\
		&minimum number of iterations	& $n_{iter,min}=3$ \\
		&maximum number of iterations   & $n_{iter,max}=6$ \\
		&maximum height			& $h_{max}=0.15 R_{\odot}$ \\
		&number of loop segments	& $n_{seg}=7$ \\
		&$\alpha$-parameter increment&$\Delta\alpha=10 R_{\odot}^{-1}$\\
\enddata
\end{deluxetable}

\begin{deluxetable}{rlrrrrrrr}
\tabletypesize{\footnotesize}
\tablecaption{Free magnetic energy and changes during X,M,C-class flares 
(\#1-36), calculated (a) with the COR-NLFFF code and 
(b) with the Wiegelmann NLFFF code (Sun et al.~2012a). The preflare
free energies are $E_{pre}^a$ nad $E_{pre}^b$, the postflare values 
are $E_{post}^a$ and $E_{post}^b$, and the changes during the flare 
are $\Delta E^a$ and $\Delta E^b$, in units of $10^{30}$ erg.}
\tablewidth{0pt}
\tablehead{
\colhead{Flare}&
\colhead{Observation}&
\colhead{GOES }&
\colhead{$E_{pre}^a$}&
\colhead{$E_{post}^a$}&
\colhead{$\Delta E^a$}&
\colhead{$E_{pre}^b$}&
\colhead{$E_{post}^b$}&
\colhead{$\Delta E^b$}\\
\colhead{\#}&
\colhead{Date}&
\colhead{class}&
\colhead{($10^{30}$ erg)}&
\colhead{($10^{30}$ erg)}&
\colhead{($10^{30}$ erg)}&
\colhead{($10^{30}$ erg)}&
\colhead{($10^{30}$ erg)}&
\colhead{($10^{30}$ erg)}}
\startdata
  1&2011-Feb-13 13:43&C4.7&  26$\pm$  6&  10$\pm$  6& -15$\pm$  9& 129$\pm$  4& 129$\pm$  4&   0$\pm$  2\\
  2&2011-Feb-13 17:27&M6.6&  65$\pm$  6&  21$\pm$  6& -44$\pm$  9& 164$\pm$  4& 142$\pm$  4& -21$\pm$  2\\
  3&2011-Feb-13 21:16&C1.1&  76$\pm$  6&  25$\pm$  6& -51$\pm$  9& 167$\pm$  4& 157$\pm$  4&  -9$\pm$  1\\
  4&2011-Feb-14  2:35&C1.6&  46$\pm$ 10&  38$\pm$ 10&  -7$\pm$ 14& 190$\pm$  4& 190$\pm$  4&   0$\pm$  3\\
  5&2011-Feb-14  4:29&C8.3&  52$\pm$ 10&  29$\pm$ 10& -23$\pm$ 14& 196$\pm$  4& 181$\pm$  4& -15$\pm$  2\\
  6&2011-Feb-14  6:50&C6.6&  75$\pm$ 10&  42$\pm$ 10& -33$\pm$ 14& 201$\pm$  4& 200$\pm$  4&   0$\pm$  2\\
  7&2011-Feb-14  8:38&C1.8&  95$\pm$ 10&  40$\pm$ 10& -54$\pm$ 14& 209$\pm$  4& 198$\pm$  4& -11$\pm$  2\\
  8&2011-Feb-14 11:50&C1.7&  44$\pm$ 10&  30$\pm$ 10& -14$\pm$ 14& 213$\pm$  4& 210$\pm$  4&  -3$\pm$  1\\
  9&2011-Feb-14 12:40&C9.4&  70$\pm$ 10&  33$\pm$ 10& -37$\pm$ 14& 219$\pm$  4& 206$\pm$  4& -13$\pm$  1\\
 10&2011-Feb-14 13:46&C7.0&  70$\pm$ 10&  46$\pm$ 10& -24$\pm$ 14& 217$\pm$  4& 208$\pm$  4&  -8$\pm$  1\\
 11&2011-Feb-14 17:20&M2.2&  60$\pm$ 10&  34$\pm$ 10& -25$\pm$ 14& 244$\pm$  4& 228$\pm$  4& -16$\pm$  3\\
 12&2011-Feb-14 19:23&C6.6&  52$\pm$ 10&  37$\pm$ 10& -14$\pm$ 14& 250$\pm$  4& 241$\pm$  4&  -9$\pm$  3\\
 13&2011-Feb-14 23:14&C1.2&  77$\pm$ 10&  44$\pm$ 10& -32$\pm$ 14& 264$\pm$  4& 259$\pm$  4&  -5$\pm$  2\\
 14&2011-Feb-14 23:39&C2.7&  62$\pm$ 10&  40$\pm$ 10& -21$\pm$ 14& 265$\pm$  4& 259$\pm$  4&  -5$\pm$  2\\
 15&2011-Feb-15  0:31&C2.7&  92$\pm$ 20&  61$\pm$ 20& -31$\pm$ 28& 259$\pm$  6& 250$\pm$  6&  -8$\pm$  4\\
 16&2011-Feb-15  1:43&X2.2&  98$\pm$ 20&  36$\pm$ 20& -62$\pm$ 28& 256$\pm$  6& 218$\pm$  6& -37$\pm$  4\\
 17&2011-Feb-15  4:26&C4.8&  61$\pm$ 20&  56$\pm$ 20&  -5$\pm$ 28& 241$\pm$  6& 230$\pm$  6& -11$\pm$  2\\
 18&2011-Feb-15 10: 1&C1.0& 150$\pm$ 20&  53$\pm$ 20& -97$\pm$ 28& 214$\pm$  6& 213$\pm$  6&   0$\pm$  3\\
 19&2011-Feb-15 14:31&C4.8& 179$\pm$ 20&  39$\pm$ 20&-139$\pm$ 28& 198$\pm$  6& 193$\pm$  6&  -4$\pm$  1\\
 20&2011-Feb-15 18: 6&C1.7& 147$\pm$ 20&  59$\pm$ 20& -88$\pm$ 28& 222$\pm$  6& 208$\pm$  6& -14$\pm$  3\\
 21&2011-Feb-15 19:30&C6.6& 122$\pm$ 20&  47$\pm$ 20& -75$\pm$ 28& 223$\pm$  6& 204$\pm$  6& -18$\pm$  3\\
 22&2011-Feb-15 22:48&C1.3&  87$\pm$ 20&  26$\pm$ 20& -61$\pm$ 28& 222$\pm$  6& 209$\pm$  6& -13$\pm$  5\\
 23&2011-Feb-16  0:57&C2.0& 103$\pm$ 34&  50$\pm$ 34& -53$\pm$ 48& 214$\pm$  6& 214$\pm$  6&   0$\pm$  3\\
 24&2011-Feb-16  1:31&M1.0& 207$\pm$ 34&  81$\pm$ 34&-126$\pm$ 48& 205$\pm$  6& 204$\pm$  6&   0$\pm$  2\\
 25&2011-Feb-16  1:55&C2.2& 136$\pm$ 34&  78$\pm$ 34& -58$\pm$ 48& 213$\pm$  6& 203$\pm$  6&  -9$\pm$  2\\
 26&2011-Feb-16  5:39&C5.9& 118$\pm$ 34&  93$\pm$ 34& -25$\pm$ 48& 219$\pm$  6& 202$\pm$  6& -16$\pm$  2\\
 27&2011-Feb-16  6:18&C2.2& 131$\pm$ 34&  70$\pm$ 34& -61$\pm$ 48& 225$\pm$  6& 197$\pm$  6& -28$\pm$  3\\
 28&2011-Feb-16  7:34&M1.1& 182$\pm$ 34&  58$\pm$ 34&-123$\pm$ 48& 205$\pm$  6& 205$\pm$  6&   0$\pm$  4\\
 29&2011-Feb-16  9: 1&C9.9& 131$\pm$ 34&  49$\pm$ 34& -82$\pm$ 48& 224$\pm$  6& 191$\pm$  6& -32$\pm$  3\\
 30&2011-Feb-16 10:25&C3.2&  61$\pm$ 34&  22$\pm$ 34& -39$\pm$ 48& 202$\pm$  6& 201$\pm$  6&   0$\pm$  3\\
 31&2011-Feb-16 11:58&C1.0&  83$\pm$ 34&  16$\pm$ 34& -66$\pm$ 48& 199$\pm$  6& 191$\pm$  6&  -7$\pm$  3\\
 32&2011-Feb-16 14:18&M1.6& 192$\pm$ 34&  34$\pm$ 34&-157$\pm$ 48& 181$\pm$  6& 169$\pm$  6& -11$\pm$  2\\
 33&2011-Feb-16 15:26&C7.7& 218$\pm$ 34&   5$\pm$ 34&-212$\pm$ 48& 180$\pm$  6& 163$\pm$  6& -16$\pm$  5\\
 34&2011-Feb-16 19:29&C1.3& 177$\pm$ 34& 135$\pm$ 34& -41$\pm$ 48& 176$\pm$  6& 170$\pm$  6&  -6$\pm$  3\\
 35&2011-Feb-16 20:11&C1.1& 179$\pm$ 34&  89$\pm$ 34& -89$\pm$ 48& 180$\pm$  6& 180$\pm$  6&   0$\pm$  3\\
 36&2011-Feb-16 21: 6&C4.2& 248$\pm$ 34& 107$\pm$ 34&-141$\pm$ 48& 159$\pm$  6& 158$\pm$  6&   0$\pm$  3\\
\enddata
\end{deluxetable}

\begin{deluxetable}{lll}
\tabletypesize{\normalsize}
\tablecaption{Energy budget for the X2.2 flare on 2011-Feb-15.
The magnetic energies labeld with the symbol $^*)$ refers to
HMI data that were subjected to pre-processing (Sun et al.~2012a;
Jiang and Feng 2013).}
\tablewidth{0pt}
\tablehead{
\colhead{Energy}&
\colhead{Energy}&
\colhead{Reference}\\
\colhead{type}&
\colhead{$E$ ($10^{30}$ erg)}&
\colhead{}}
\startdata
\underbar{Energy Input :}		&	      &	                      \\
Total nonpotential energy $E_{N}$	& $1210^*$    & Sun et al.~(2012a)    \\
					& $983-1090^*$& Jiang and Feng (2013) \\
                                        & $492\pm158$ & Vemareddy et al.~(2012b)\\
                                        & $ 857$      & (This study)          \\
Total potential energy $E_{P}$		& $ 970^*$    & Sun et al.~(2012a)    \\
					& $818-897^*$ & Jiang and Feng (2013) \\
                                        & $243\pm78$  & Vemareddy et al.~(2012b)\\
                                        & $ 760$      & (This study)          \\
Free magnetic energy $E_{free}=E_{N}-E_P$& $240$      & Sun et al.~92012a)    \\
					& $165-193$   & Jiang and Feng (2013) \\
                                        & $98\pm20$   & (This study)          \\
Decrease of free energy $E_{post}-E_{pre}$ & $37\pm 4$& Sun et al.~(2012a)    \\
                                        & $62\pm 28$& (This study)            \\
Minimum Current Corona (MCC) energy     & $168$       & Tarr et al.(2013)     \\
Energy drop from virial theorem         & $105\pm 104$& Vemareddy et al.~(2012b) \\
Energy drop from virial theorem         & $84$        & Tziotziou et al.~( 2013) \\
\hline
\underbar{Energy output :}		&		&		        \\ 
Nonthermal energy RHESSI (25-50 keV)    & $54$        & Sun et al.~(2012a)     \\
Thermal energy AIA/SDO 		        & $20$        & Aschwanden et al.~(2013a) \\    
Radiated energy in Ly alpha             & $1.0$       & Milligan et al.~(2012)   \\
Radiatied in free-free continuum        & $0.8$       & Milligan et al.~(2012)   \\
Radiated in GOES X-rays 1-8 A           & $0.5$       & Milligan et al.~(2012)   \\
Radiated in Lyman continuum             & $0.4$       & Milligan et al.~(2012)   \\
Radiated in He I 304 A                  & $0.3$       & Milligan et al.~(2012)   \\
Radiated in He I cont                   & $0.04$      & Milligan et al.~(2012)   \\
Radiated in He II cont                  & $0.01$      & Milligan et al.~(2012)   \\
Helioseismic energy in 3-4 mHz band     & $0.0018$    & Alvarado-Gomez et al.~(2012)\\
\enddata
\end{deluxetable}

\begin{deluxetable}{rrrrrrrrr}
\tabletypesize{\normalsize}
\tablecaption{Energy budget of X,M-class flares during 2011 Feb 12-16:
$\Delta E_{free}^a$ = change of free energy calculated with COR-NLFFF code in this study;
$\Delta E_{free}^b$ = change of free energy calculated with the photospheric
(Wiegelmann) NLFFF code (Sun et al.~2012a);
$\Delta E_{MCC}$ = energy change from Minimum Current Corona model (Tarr et al.~2013);
$\Delta E_{VT}^c$ = energy change from virial theorem (Vemareddy et al.~2012b);
$\Delta E_{VT}^d$ = energy change from virial theorem (Tziotziou et al.~2013);
$E_{th}$ = thermal energy from AIA/SDO DEM analysis (Aschwanden et al.~2013a).}
\tablewidth{0pt}
\tablehead{
\colhead{Flare}&
\colhead{Observation}&
\colhead{GOES }&
\colhead{$\Delta E_{free}^a$}&
\colhead{$\Delta E_{free}^b$}&
\colhead{$\Delta E_{MCC}$}&
\colhead{$\Delta E_{VT}^c$}&
\colhead{$\Delta E_{VT}^d$}&
\colhead{$\Delta E_{th}$}\\
\colhead{\#}&
\colhead{Date}&
\colhead{class}&
\colhead{($10^{32}$ erg)}&
\colhead{($10^{32}$ erg)}&
\colhead{($10^{32}$ erg)}&
\colhead{($10^{32}$ erg)}&
\colhead{($10^{32}$ erg)}&
\colhead{($10^{32}$ erg)}}
\startdata
  2&2011-Feb-13 17:27 &M6.6 & -44$\pm$  9  &-21$\pm$2 &  -4 & -43$\pm$ 17 &-10 & 25 \\
 11&2011-Feb-14 17:20 &M2.2 & -62$\pm$ 28  &-16$\pm$3 &  -3 &             &-49 &    \\
 16&2011-Feb-15  1:43 &X2.2 & -68$\pm$ 20  &-37$\pm$4 &-168 &-105$\pm$104 &-84 & 20 \\
 24&2011-Feb-16  1:31 &M1.0 &-126$\pm$ 48  &  0$\pm$2 &     &             &    &  3 \\
 28&2011-Feb-16  7:34 &M1.1 &-123$\pm$ 48  &  0$\pm$4 &     &             &    &  2 \\
 32&2011-Feb-16 14:18 &M1.6 &-157$\pm$ 48  &-11$\pm$2 &     &             &-65 &  5 \\
\enddata
\end{deluxetable}

\clearpage


\begin{figure}
\plotone{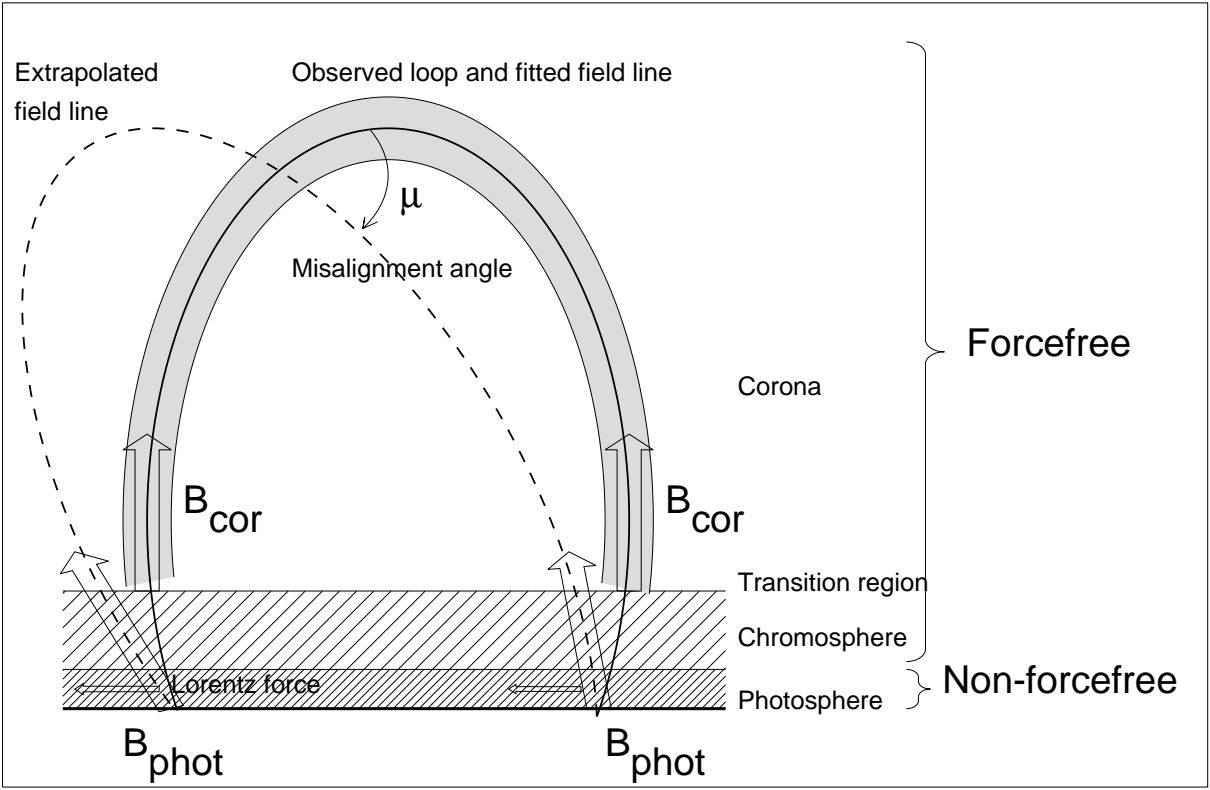}
\caption{The diagram shows the misalignment between a magnetic field line
(dashed curve) that is extrapolated from the magnetic field vectors
${\bf B}_{phot}$ from the non-forcefree photosphere, and a 
magnetic field line (solid curve) that is obtained from forward-fitting
of a nonlinear forcefree field model ${\bf B}_{cor}$ to an observed loop 
geometry (grey color), quantified by a misalignment angle $\mu$.}
\end{figure} 

\begin{figure}
\plotone{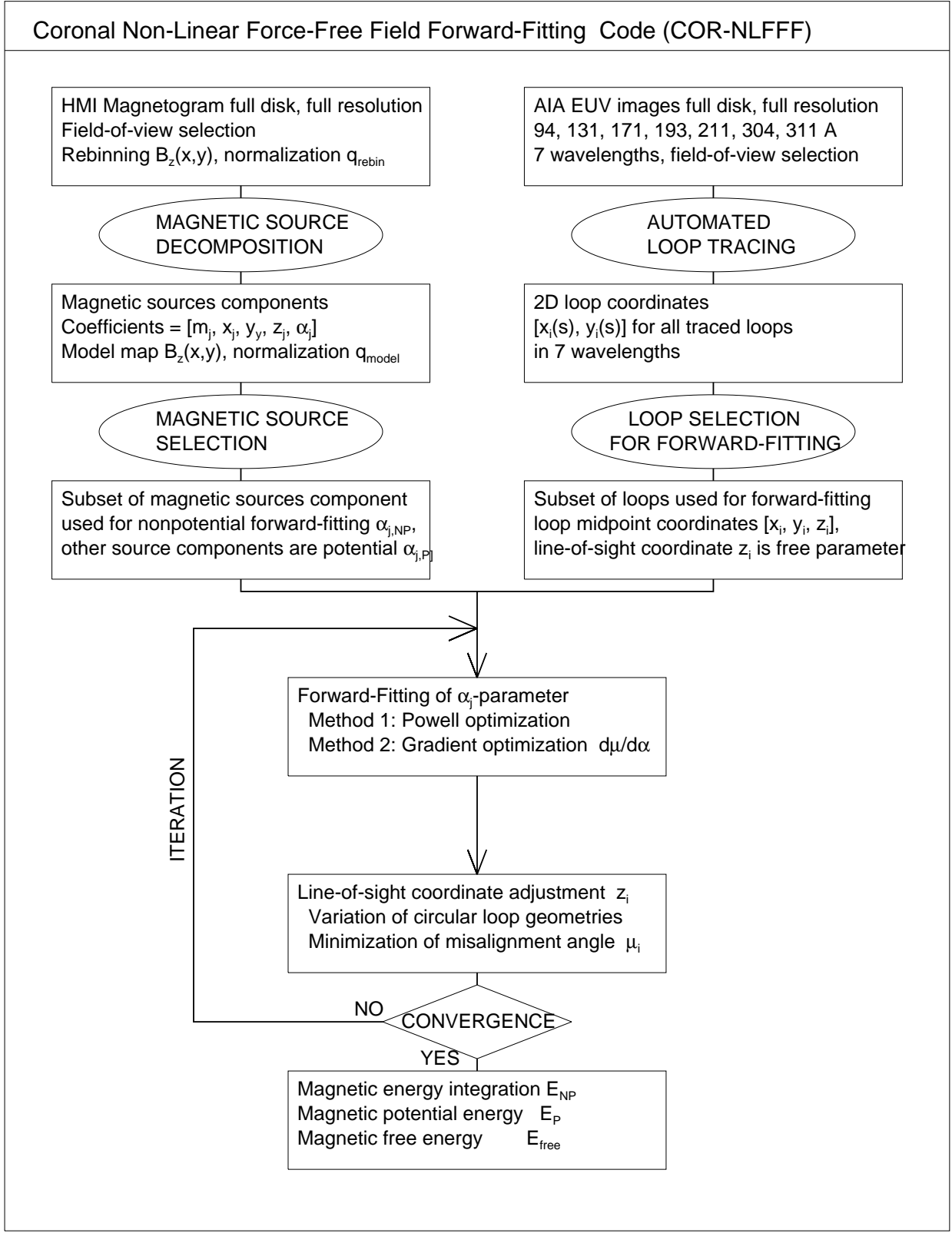}
\caption{Flow chart of the COR-NLFFF code, which includes processing
of the magnetic data (top left), of the EUV image data (top right),
and forward-fitting using both data sets (bottom half). See Section
2 for a theoretical description, and Section 3 and Appendix A for 
parametric tests.}
\end{figure}

\begin{figure}
\plotone{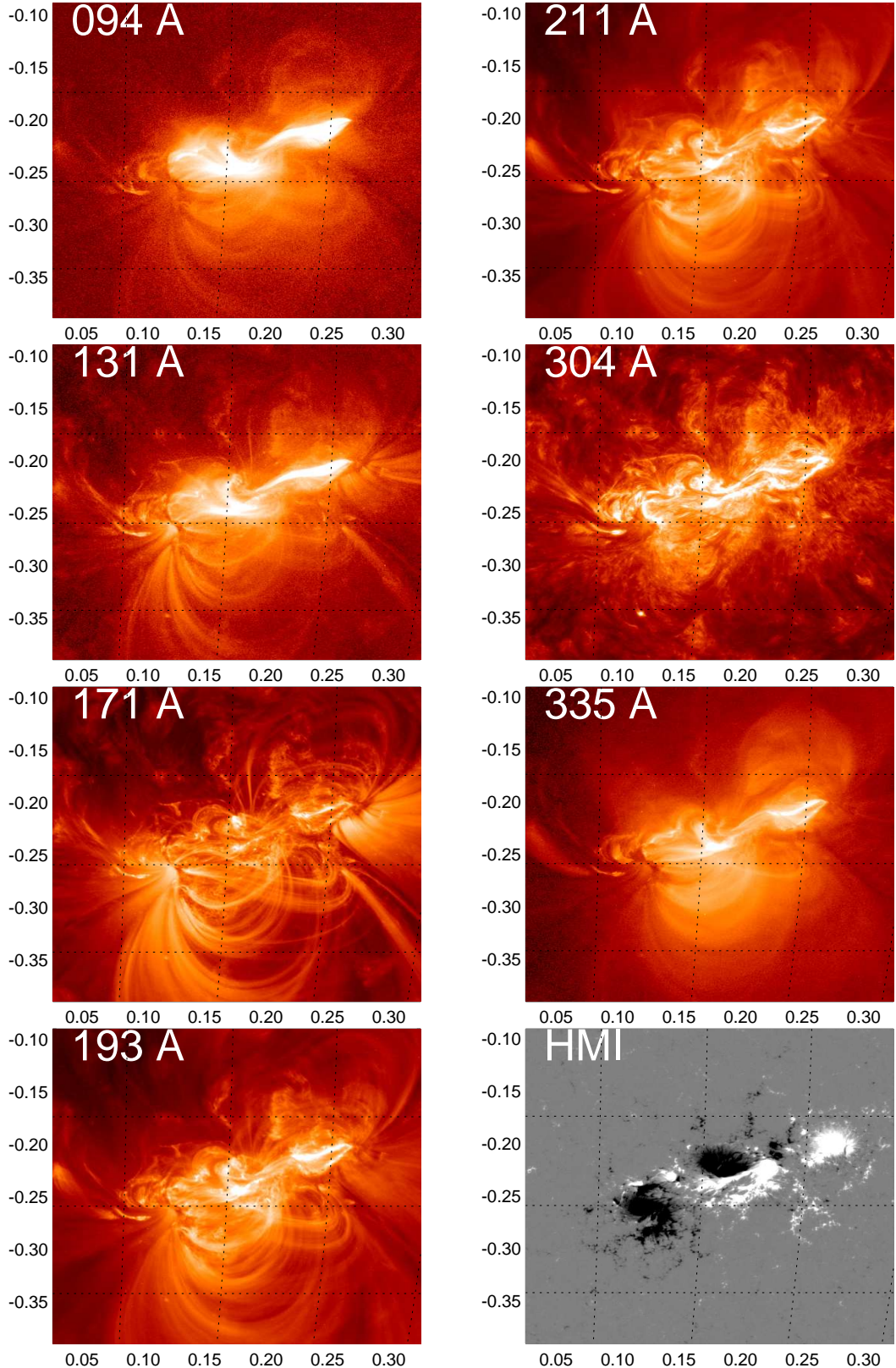}
\caption{A set of 7 EUV images observed with AIA/SDO between
2011-02-15 00:00:00 UT and 00:00:12 UT, in the wavelengths
of 94, 131, 171, 193, 211, 304, 335 \ang , rendered on a
logarithmic color scale. The co-spatial and cotemporanous
HMI/SDO magnetogram (bottom right) was observed on 2011-02-14 
23:58:57 UT, rendered in greyscale. The field-of-view of all
images is $x=[0.03-0.33]$ solar radii in EW direction and
$y=[-0.39,-0.09]$ solar radii in NS direction, corresponding to
a range of 210 Mm.}
\end{figure}

\begin{figure}
\plotone{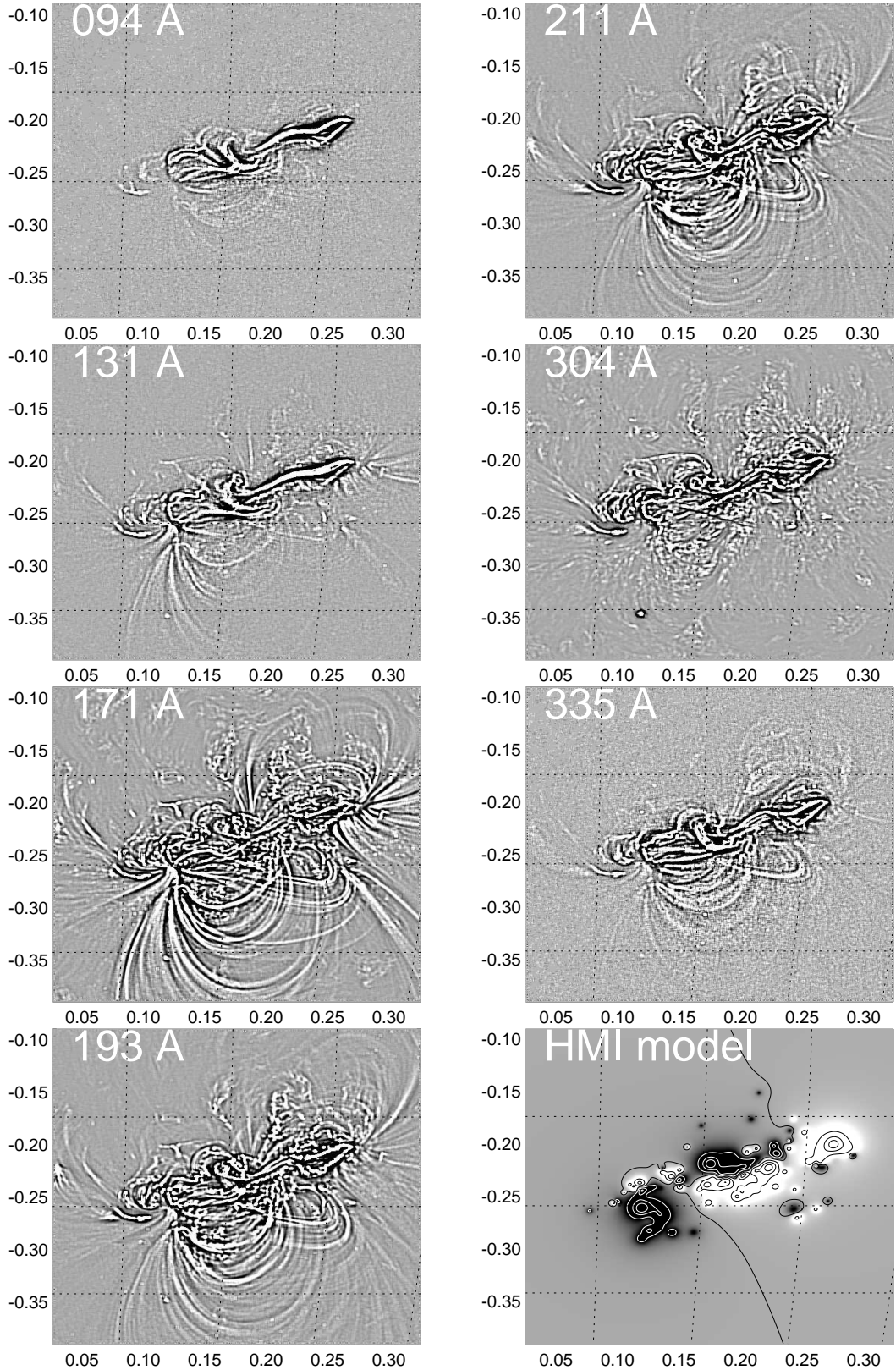}
\caption{The same dataset of AIA images shown in Fig.~3 is bandpass-filtered
with a lowpass filter boxcar of $nsm_1=5$ pixels and a highpass filter
boxcar of $nsm_2=7$ pixels. The HMI magnetogram (bottom right) is decomposed
into 100 Gaussian-like magnetic source components and superimposed to a
LOS model map that is parameterized with $4 \times 100$ parameters.}
\end{figure}

\begin{figure}
\plotone{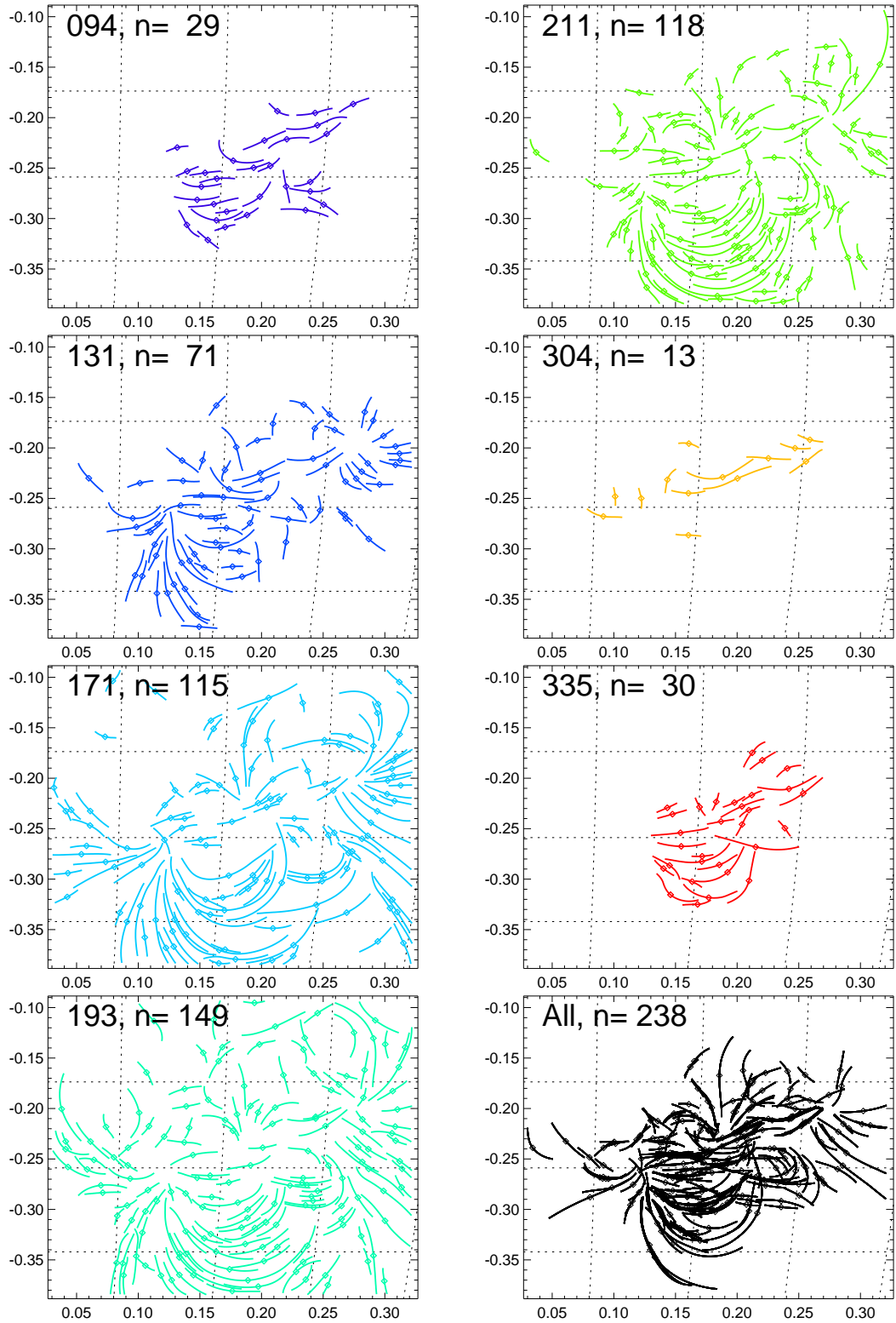}
\caption{Automated loop tracing of the 7 bandpass-filtered AIA images 
(shown in Fig.~4) with the code OCCULT-2. Each wavelength is represented
with a different color, and a composite of all wavelengths (containing a limit
of 50 loop structures per wavelength) is shown in the
bottom right panel. The number $n_{loop}$ of loop segments detected above a noise
threshold and with a minimum length of 30 pixels is indicated.}
\end{figure}

\begin{figure}
\plotone{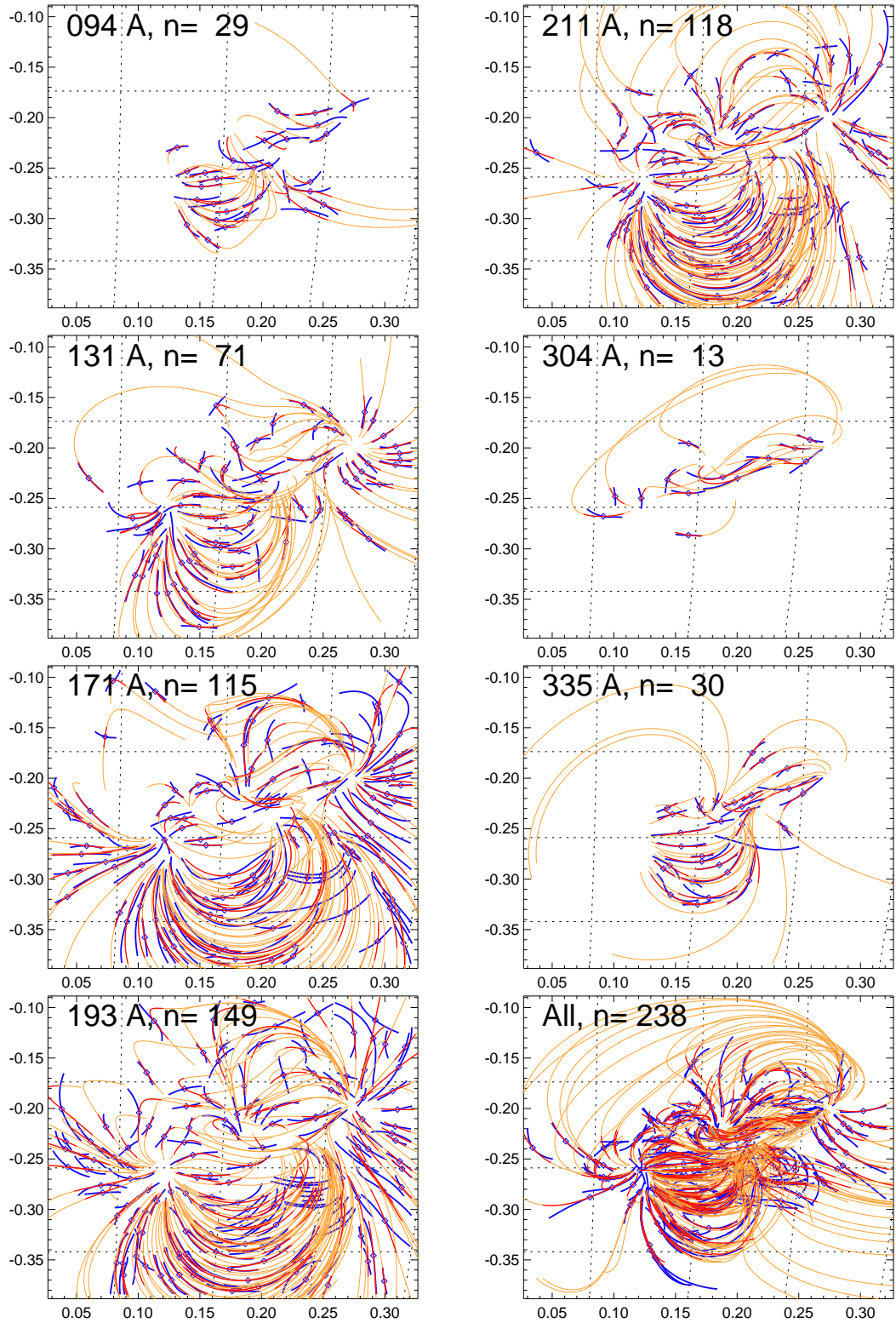}
\caption{Forward-fitting of NLFFF magnetic field approximation (orange curves)
to the automatically traced loops (blue curves) in each wavelength filter.
The locations of the theoretical field lines where chosen at the intersection
of the midpoints (diamonds) of traced loops, and a field line segment of equal 
length as the traced loop segment is shown with a red curve. The bottom right
panel shows a simultaneous fit to a subset of $n=238$ loops synthesized from 
the 7 AIA filters.}
\end{figure}

\begin{figure}
\plotone{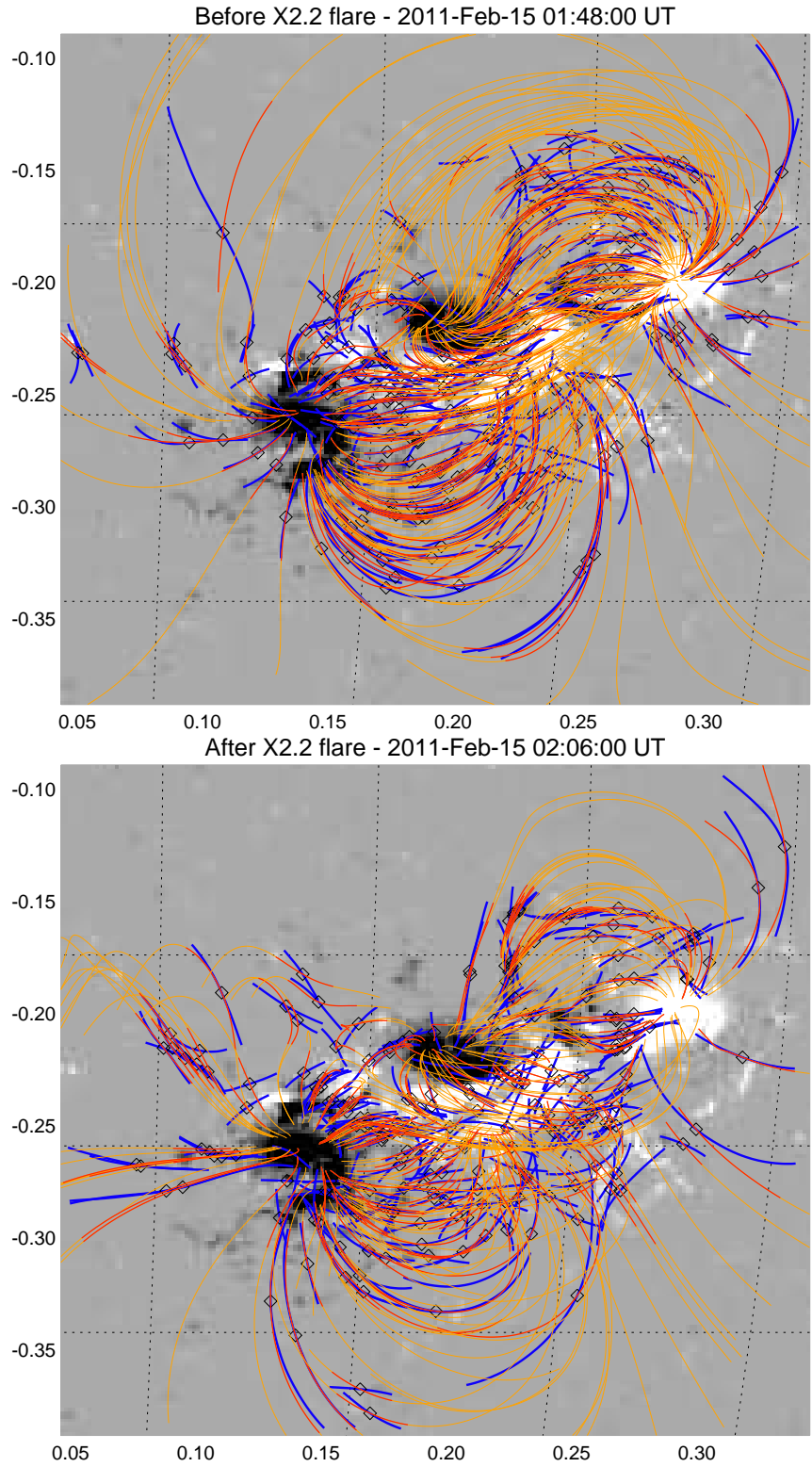}
\caption{Best-fit solution of nonlinear force-free field (orange curves)
overlaid on the automatically traced coronal loops (blue curves),
tailored to the same loop length segments (red curves), before the
X2.2 flare on 2011-Feb-15, 01:48 UT (top panel), and after the
flare peak on 2011-Feb-15, 02:06 UT (bottom panel), overlaid on the 
line-of-sight HMI magnetogram. The time range covers the interval
of the largest energy decrease of free magnetic energy during the flare.}
\end{figure}

\begin{figure}
\plotone{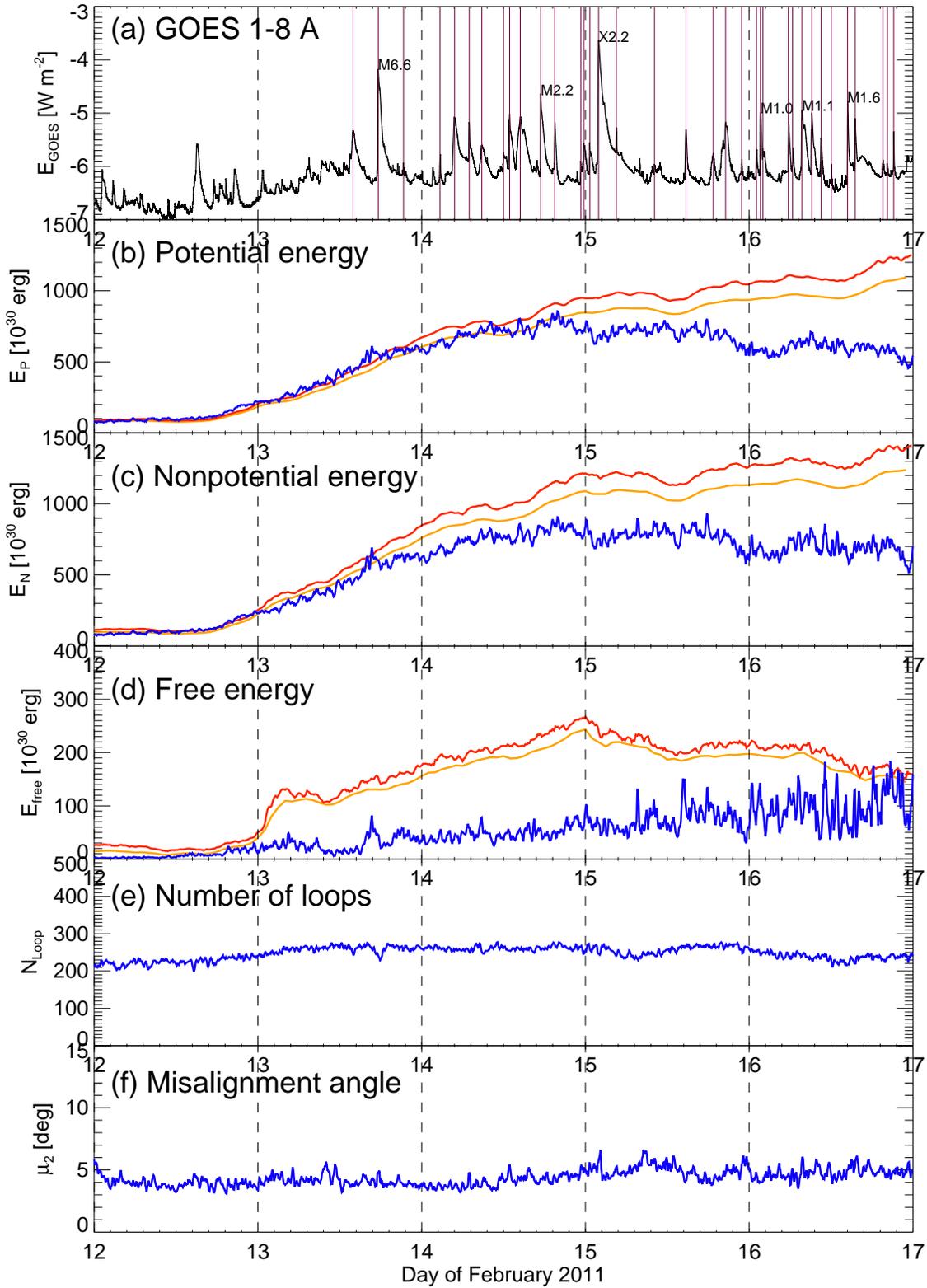}
\caption{Time evolution of magnetic energies of AR 11158
during 2011 Feb 12 to 17: (a) GOES 1-8 \ang\ flux, with GOES C-, M-, 
and X-class flares indicated with purple vertical lines; 
(b) Potential field energy $E_P$; 
(c) Nonpotential energy $E_N$;
(d) Free energy $E_{free}=E_{N}-E_P$; 
(e) The number of fitted loops $N_{loop}$; 
(f): the 2D misalignment angle $\mu_2$ of the best fit. The color code 
indicates forward-fitting of traced loops with the COR-NLFFF code in 
6-min time intervals (blue), the Wiegelmann NLFFF code in 12-min intervals 
(red), and 1-hr intervals (orange; Sun et al.~2012a).}
\end{figure}

\begin{figure}
\plotone{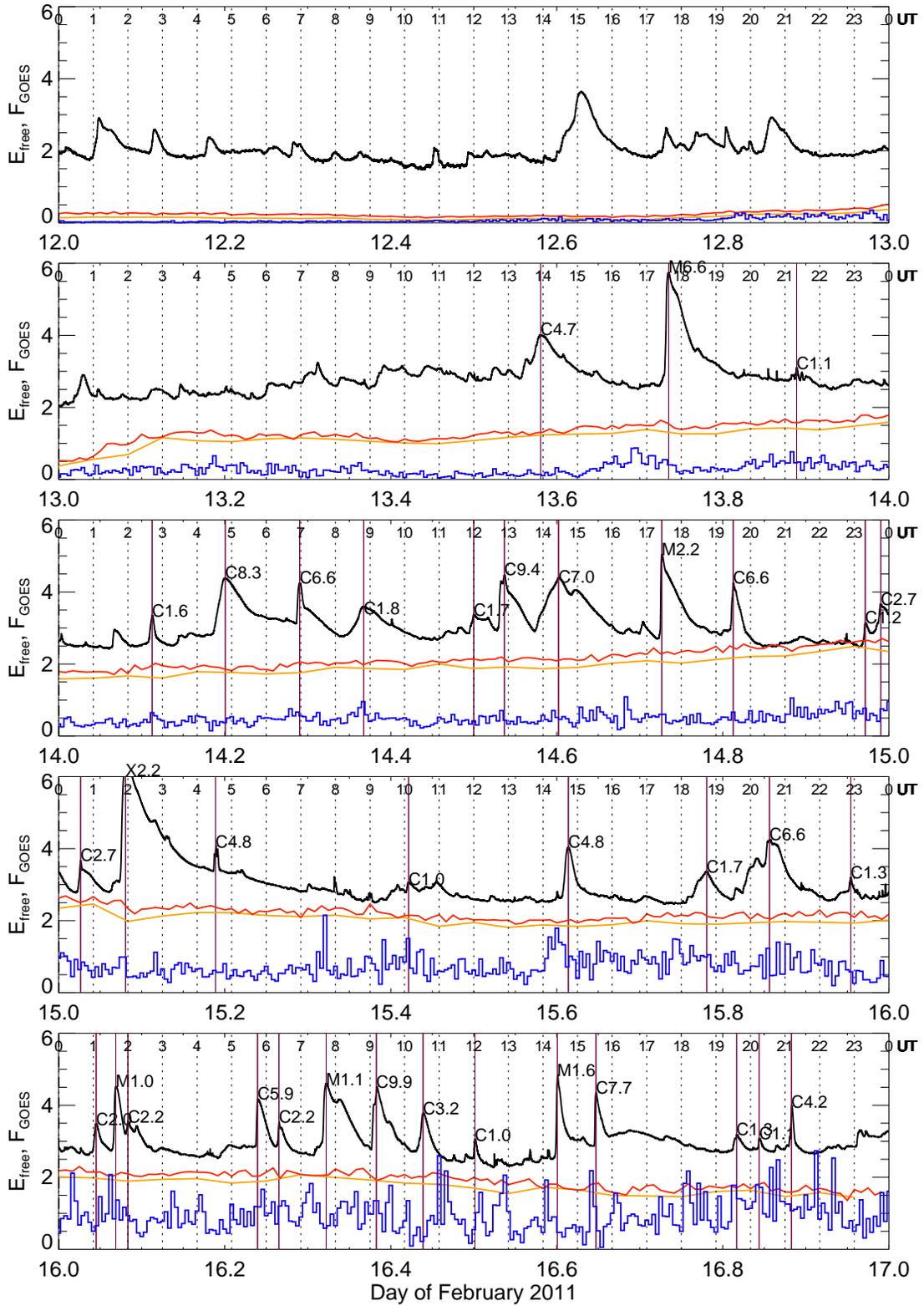}
\caption{Expanded time profiles of the GOES 1-8 \ang\ flux (black,
arbitrary units), the free energies computed with the Wiegelmann code
in 12-min intervals (red) and 1-hr intervals (orange; Sun et al.~2012a),
with forward-fitting of automatically traced loops in 6-min intervals
(blue). The times of 36 GOES C-,M-, and X-class flares occurring in
AR 11158 are indicated with vertical purple lines and labeled with 
the GOES class. Each panel represents a consecutive day from 
2011 Feb 12 to 17.}
\end{figure}

\begin{figure}
\plotone{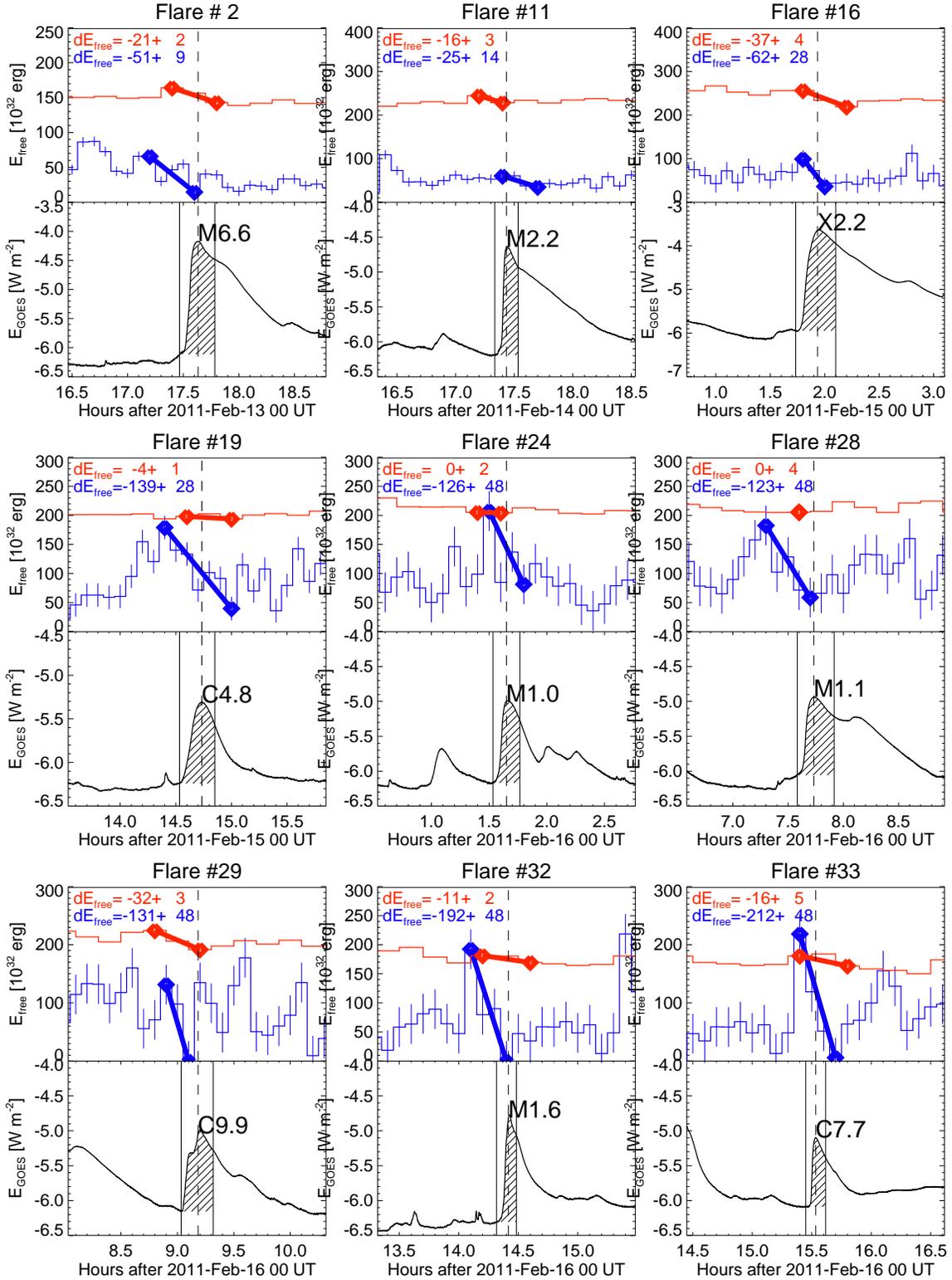}
\caption{The change in free magnetic energy is shown during 9 flares
(shown in 9 different panels), including the GOES 1-8 \ang\ flux
(black), bracketed between start ($t_{start}$) and end time ($t_{end}$)
(hatched between vertical lines), and peaking at $t_{peak}$ (dashed
vertical line), the free energy $E_{free}$ computed with forward-fitting
to coronal loops (blue histrograms), and with the Wiegelmann
NLFFF code (red histrograms). The decrease of the free energy is measured
between the maximum free energy in the preflare interval [$t_{start}-0.3$
hr, $t_{peak}$] and the minimum free energy in the flare decay time
interval $[t_{peak},t_{peak}+0.3]$ hr (indicated with diamonds and thick
solid blue and red curve).}
\end{figure}

\begin{figure}
\plotone{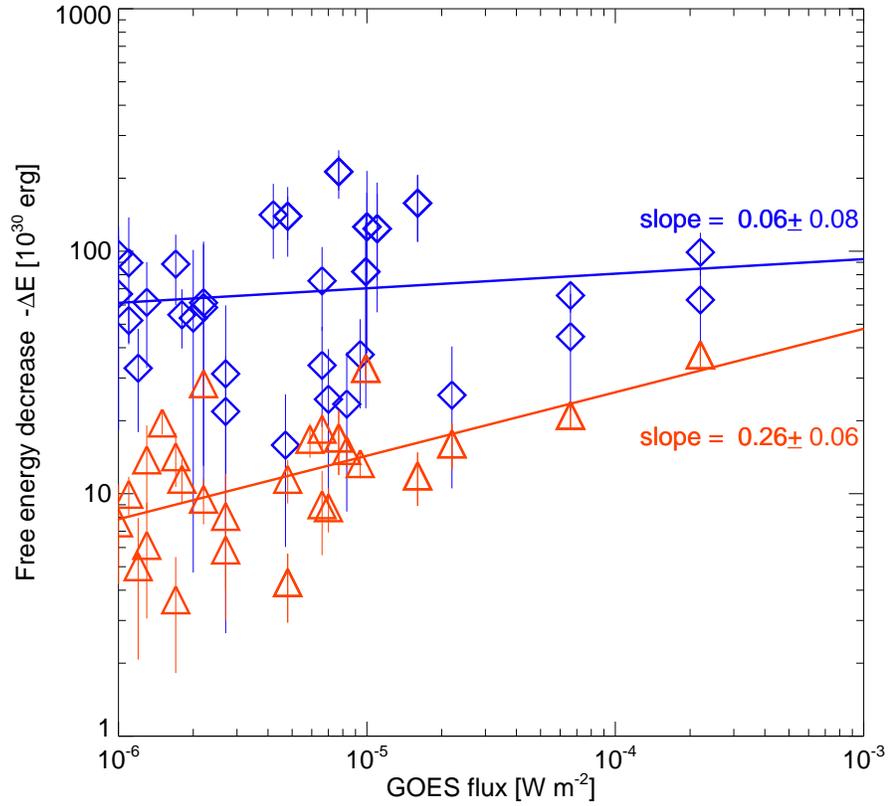}
\caption{Scatterplot of the decrease of free magnetic energy
$(-\Delta E_{free})$ with the GOES flux [W m$^{-2}$] for all analyzed
flare events with significant decreases. The energy drops calculated
with the Wiegelmann NLFFF code are indicated with red triangles, and
those with the COR-NLFFF code with blue diamonds. Linear regression
fits are indicated. Note the order of magnitude difference in energy
decreases between the two codes.}
\end{figure}

\begin{figure}
\plotone{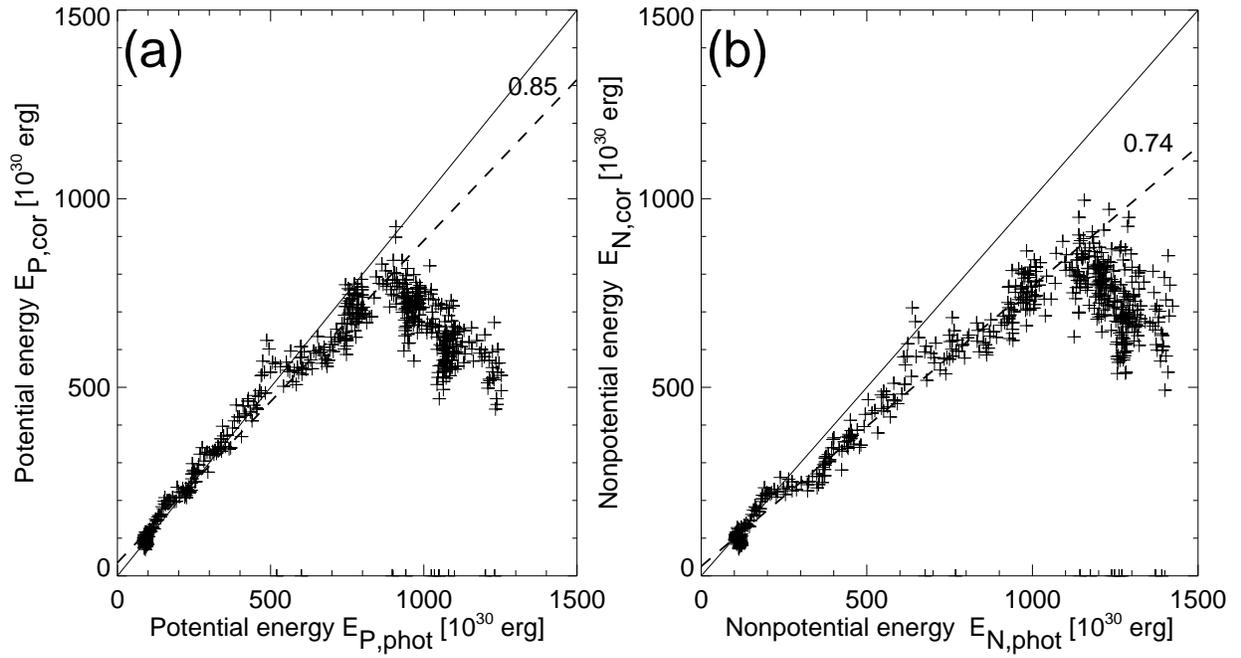}
\caption{Correlation of the potential energy $E_P$ (left panel) and the
free energy $E_{free}=E_{N}-E_P$ (right panel) for the two codes,
i.e., the Wiegelmann NLFFF code (y-axis) and the COR-NLFFF code using
forward-fitting to coronal loops (x-axis).}
\end{figure}

\begin{figure}
\plotone{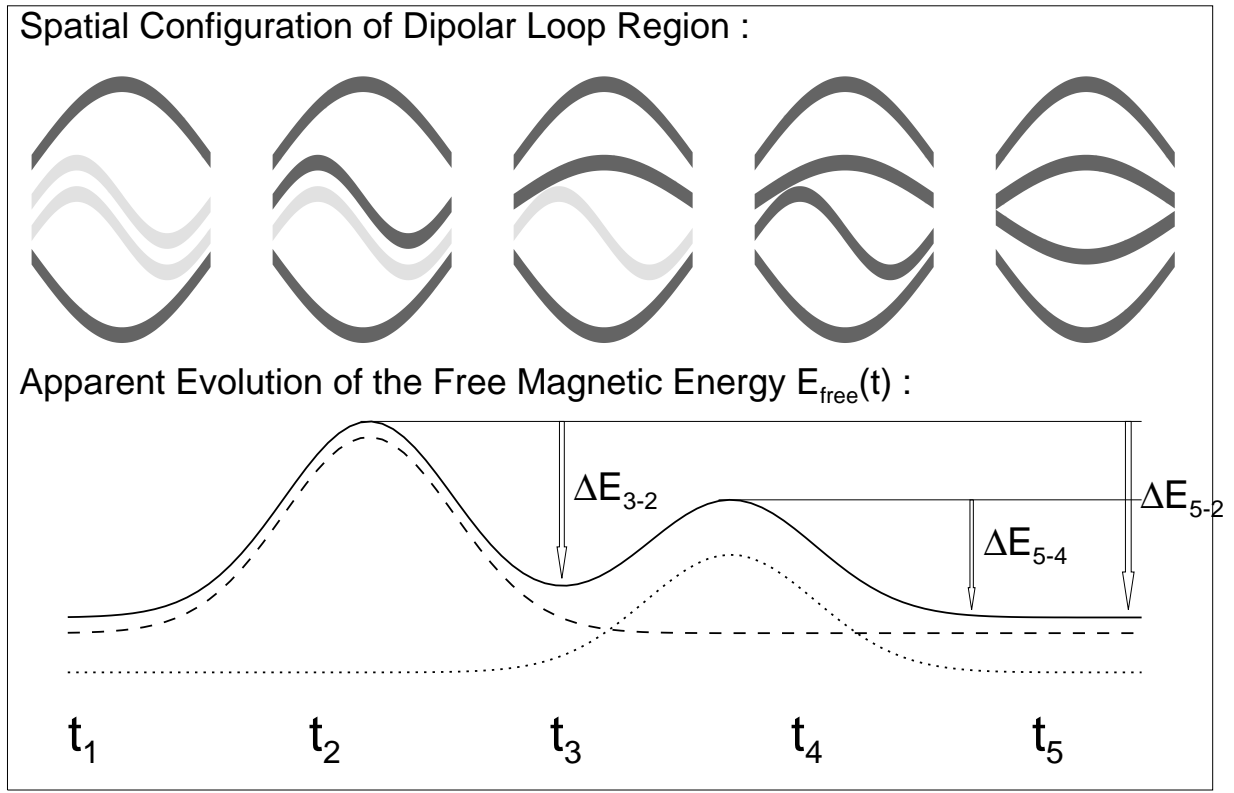}
\caption{Schematic diagram of the spatial loop configuration (top panel)
and evolution of free magnetic energy $E_{free}(t)$ during a flare.
Mostly potential loops are visible at the beginning of a flare ($t_1$),
while a first sigmoid is illuminated at $t_2$, which relaxes to a potential
loop at atim $t_3$. A second sigmoid is illuminated at time $t_4$, which
relaxes to a potential loop at time $t_5$. The total energy difference
before and after the flare, $\Delta E_{5-2}$, is a lower limit to the
sum of all sequential energy releases $\Delta E_{3-2}$ and $\Delta E_{5-4}$,
and thus underestimates the total dissipated magnetic energy.}
\end{figure}

\begin{figure}
\plotone{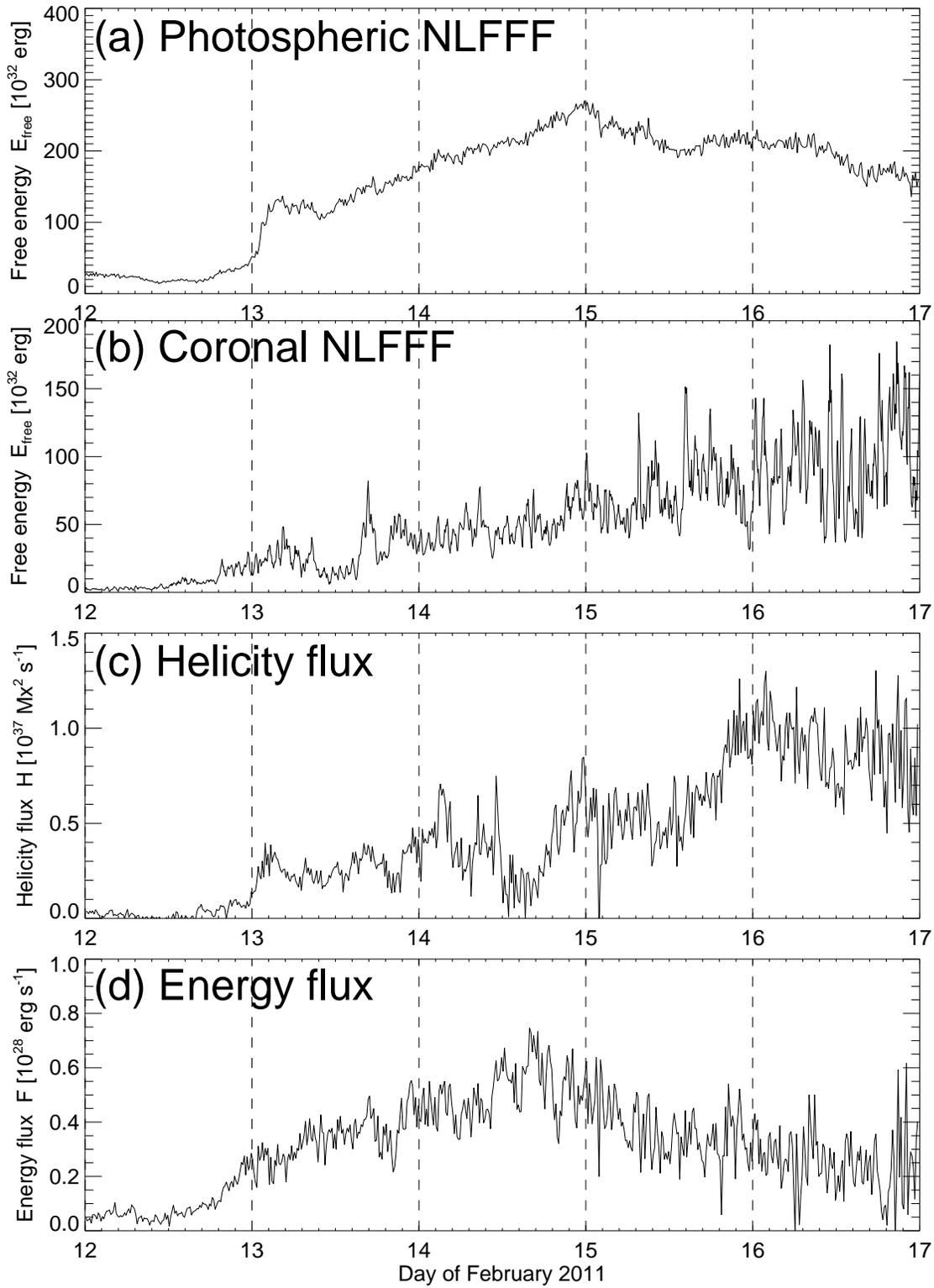}
\caption{Comparison of the free energy calculated with the photospheric 
NLFFF code (a), with the coronal NLFFF code (b), the helicity flux (c),
and the energy flux (d). Note that the photospheric NLFFF code produces
very little variability, while the coronal NLFFF code has a much higher 
degree of variability.}
\end{figure}

\begin{figure}
\plotone{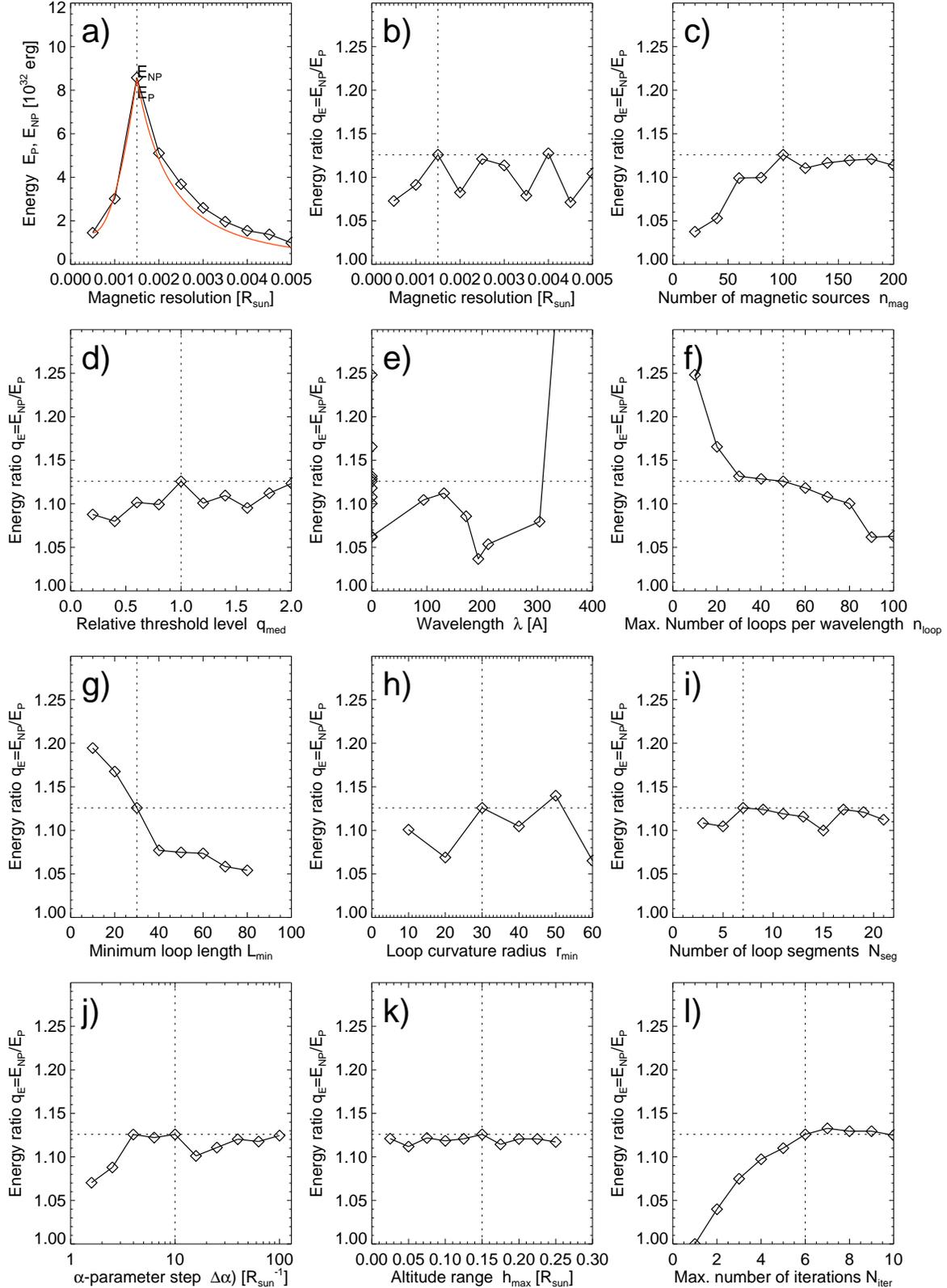}
\caption{(a) The nonpotential energy $E_{N}$ and potential energy $E_P$ 
as a function of the magnetic resolution $\Delta x_{mag}$, along with a
theoretical model (red curves) explained in Section 3.1. 
The panels (b) through (l) contain 11 parametric studies of the
energy ratio $q_E=E_{N}/E_P$ (diamonds) as a function 
of 11 control parameters varied over some range.
The vertical dotted lines indicate the chosen default values, and 
the horizontal dotted lines indicate the energy ratio at 
the default value. See full description in Appendix A.}
\end{figure}

\begin{figure}
\plotone{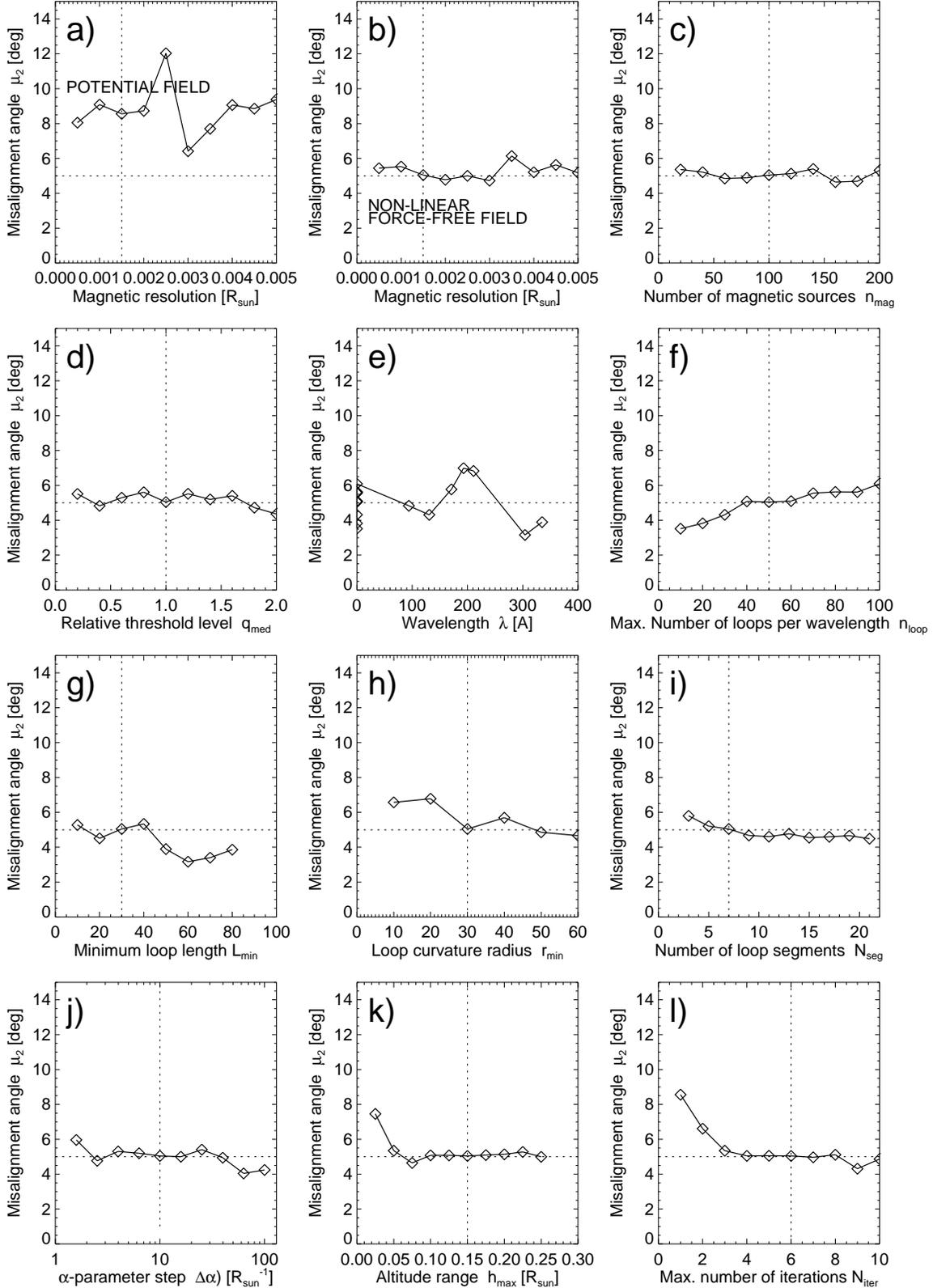}
\caption{(a) Misalignment angle $\mu_2$ for the potential field as
a function of the magnetic resolution from 1 to 10 HMI pixels.
The panels (b) through (l) contain 11 parametric studies of the
misalignment angle $\mu_2$ (diamonds) as a function 
of 11 control parameters varied over the same range as in Fig.~11.
The vertical dotted lines indicate the chosen default values, and 
the horizontal dotted lines indicate the misalignment angle at 
the default value.}
\end{figure}

\begin{figure}
\plotone{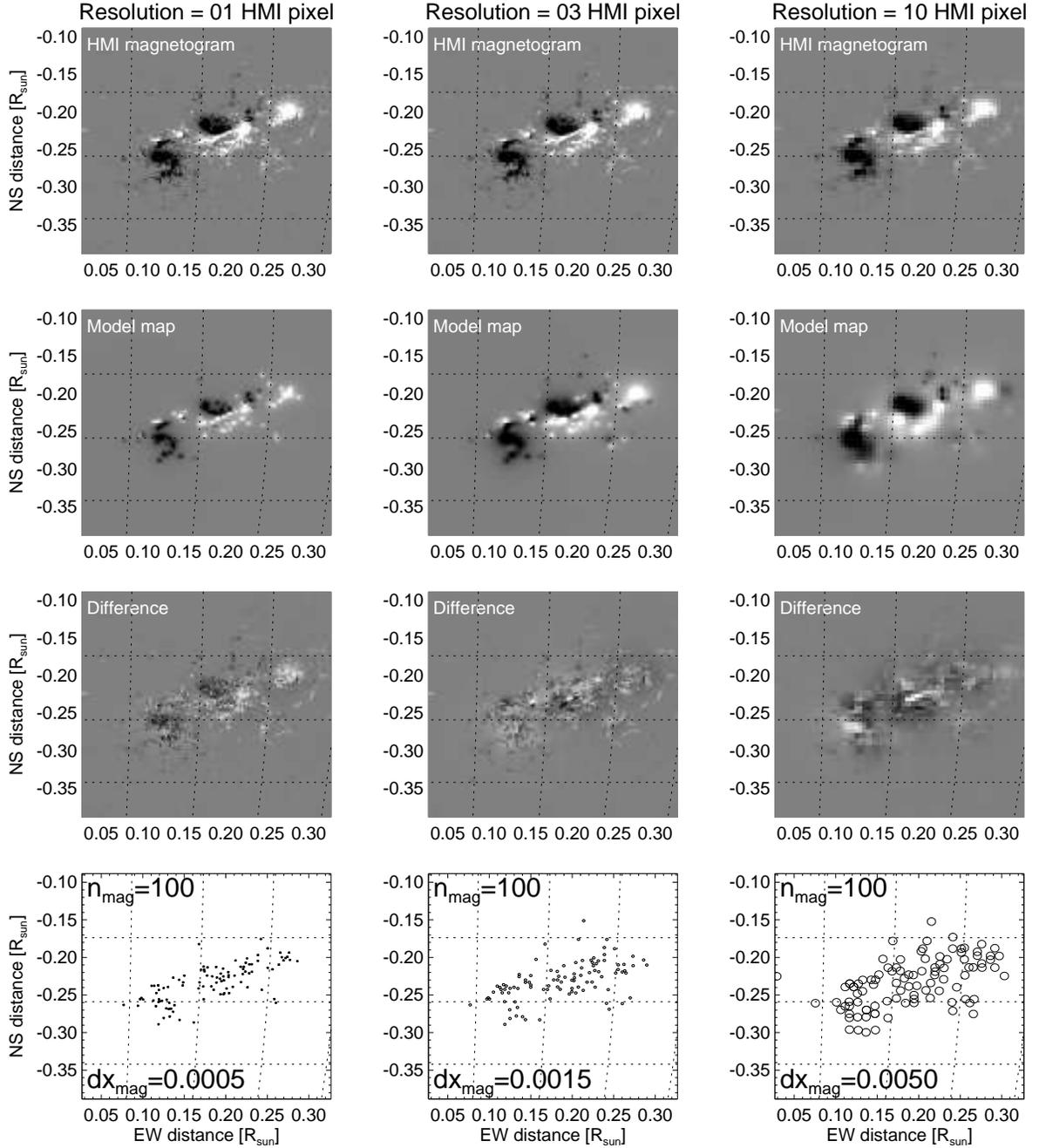}
\caption{Original HMI magnetogram $B_z(x,y)$ with full resolution
(top left), rebinned with 3 HMI pixels (top middle), and rebinned
with 10 HMI pixels (top right). 
The magnetograms are decomposed with the three different 
resolutions into 100 magnetic sources that form the model map
(second row). The difference between the observed magnetograms and
the model maps are shown in the third row with identical greyscale.
The location and radii are shown in the bottom row.}
\end{figure}

\end{document}